\documentclass[useAMS,usenatbib,usegraphicx]{mn2e}
\usepackage{amsmath,amsfonts,amssymb,url,pslatex,color}
\usepackage[T1]{fontenc}
\usepackage[utf8]{inputenc}
\DeclareUnicodeCharacter{0307}{\'{z}}
\newcommand\muarae{{HD\,160691}}
\newcommand\muArae{{$\mu$~Arae}}
\newcommand\hd{{HD\,160691}}     % Pepe2007        KG2007
\newcommand\araeD{{Dulcinea}}    % 9.6386  9.638   9.6369 (0.005)
\newcommand\araeE{{Rocinante}}   % 310.55  308.4   307.475 (1.5)
\newcommand\araeB{{Quijote}}     % 643.25  644.9   646.48 (1.5)
\newcommand\araeC{{Sancho}}      % 4205.8  4019    4472.9 (1300)
\def\hr8799{{HR\,8799}}
\def\gj691{{GJ~691}}
\newcommand{\epk}{JD\,2450915.29}
\newcommand{\epl}{JD\,2457273.2878}

\newcommand{\Chinu}{\chi^2_{\nu}}
\newcommand{\ms}{m\,s$^{-1}$}
\newcommand{\Nm}{N_{\rm RV}}

\newcommand{\megno}{MEGNO}
\newcommand{\emcee}{{\tt emcee}}
\newcommand\Done{${\cal D}_1$}
\newcommand\Dtwo{${\cal D}_2$}
\newcommand\Dthree{${\cal D}_3$}
\newcommand{\lnL}{\ln {\cal L}}

\newcommand{\Ym}{\langle Y \rangle}
\newcommand{\Npl}{N_{\rm pl}}
\def\gaia{{\em Gaia}}
\def\hipparcos{{\em  Hipparcos}}
\def\coralie{CORALIE}
\def\alma{ALMA}
\def\rem{REM}
\def\gost{{\tt GOST}}
\def\ics{ICs}
\def\ucles{UCLES}
\def\harps{HARPS}
\def\htof{{\tt  htof}}
\def\iad{IAD}

\def\hst{{\em HST}}
\def\aic{AIC}
\def\gbs{GBS}
\def\fgs{FGS}
\def\mle{MLE}

\def\rebound{{\tt REBOUND}}

\def\bic{BIC}
\def\idm#1{{\mbox{\scriptsize #1}}}

\def\vec#1{{\pmb #1}}

\def\Y{\langle Y \rangle}

\def\mean#1{\langle{}#1{}\rangle}

\newcommand{\au}{\mbox{au}} % IAU convention
\newcommand{\msun}{\mbox{m}_{\odot}}
\newcommand{\mJ}{\mbox{m}_{\idm{Jup}}}
\newcommand{\Mmean}{\mathcal{M}}

\newcommand{\ca}{a_{\idm{2}}}
\newcommand{\ab}{a_{\idm{3}}}
\newcommand{\ac}{a_{\idm{4}}}

\newcommand{\ee}{e_{\idm{2}}}
\newcommand{\eb}{e_{\idm{3}}}
\newcommand{\ec}{e_{\idm{4}}}
\newcommand{\Pb}{P_{\idm{3}}}
\newcommand{\Pc}{P_{\idm{4}}}
\newcommand{\Pd}{P_{\idm{1}}}
\newcommand{\Pe}{P_{\idm{2}}}

\newcommand{\me}{m_{\idm{2}}}
\def\deg{{\rm o}}
\def\idm#1{{\mbox{\scriptsize #1}}}
\newcommand\Chi{{\chi^2_\nu}}
\newcommand{\tepoch}{t_0}

\newcount\exno
\definecolor{myred}{rgb}{0.25,0.02,0.02}
\definecolor{myblue}{rgb}{0.2,0.0,0.7} 
\definecolor{mybrown}{rgb}{0.9,0.4,0.3}
\newcommand\ednote[1]{{\global\advance\exno by 1\
    \color{myblue}$\spadesuit$({\bfseries\the\exno}).
\color{myblue}\bfseries\em #1}}

\newcommand\hide[1]{}
\title[The $\mu$ Arae planetary system]
{The orbital architecture and stability of the $\mu$~Arae planetary system}
\author[K.~Go\'zdziewski]%
{K. Go\'zdziewski \\
Institute of Astronomy, Faculty of Physics, Astronomy and Informatics,
Nicolaus Copernicus University, Grudzi\c{a}dzka 5, 87-100 Toru\'n, Poland
}
\begin{document}
%_______________________________________________________________________________
%
\maketitle
\begin{abstract}
We {re-analyze} the global orbital architecture and dynamical stability of the \muarae{} planetary system. We have updated the best-fit elements and minimal masses of the planets based on literature {precision radial velocity (RV) measurements, now spanning 15~years. This is twice the RVs interval used for the first characterization of the system in 2006.} It consists of a Saturn- and two Jupiter-mass planets in low-eccentric orbits resembling the Earth-Mars-Jupiter configuration in the Solar system, as well as the close-in {warm} Neptune with a mass of $\simeq 14$~Earth masses. Here, we constrain this early solution with the outermost period to be accurate to {one month}. The best-fit {Newtonian model} is characterized by moderate eccentricities of the most massive planets below $0.1$ with small uncertainties $\simeq 0.02$. It is close but meaningfully separated from the 2e:1b mean motion resonance of the Saturn-Jupiter-like pair, but may be close to weak three-body MMRs. The system appears rigorously stable over a safely wide region of parameter space covering uncertainties of several $\sigma$. {The system stability is robust to a five-fold increase in the minimal masses, consistent with a wide range of inclinations, from $\simeq 20^{\circ}$ to $90^{\circ}$. This means that all planetary masses are safely below the brown dwarf mass limit. We found a weak statistical indication of the likely system inclination $I \simeq$ $20^{\circ}$--$30^{\circ}$.} With the well constrained orbital solution, we also investigate the structure of hypothetical debris disks, which are analogs of the Main Belt and Kuiper Belt, and may naturally occur in this system. 
\end{abstract}
\begin{keywords}
  celestial mechanics - planets and satellites: dynamical evolution and stability - stars: individual: HD~160691 - methods: data analysis - methods: observational - techniques: radial velocities
\end{keywords}
%_______________________________________________________________________________
%
\date{Accepted .... Received ...; in original form ...}
\pagerange{\pageref{firstpage}--\pageref{lastpage}} \pubyear{2022}
\maketitle
\label{firstpage}
%_______________________________________________________________________________
%
\section{Introduction}
%_______________________________________________________________________________
%
\hd{} (\muArae{}, \gj691{}) is a bright ($V=5.15$~mag) Sun-like, main-sequence G3IV-V dwarf monitored in a few long-term, precision radial velocity (RV) surveys.   The Anglo-Australian Telescope team (AAT, \ucles{} spectrometer) discovered its Jupiter-mass companion \hd{}b in about of 630~days orbit \citep{Butler2001}, and \cite{Jones2002a} found a linear trend in the RV data indicating a second, more distant planet. The star was also observed in the Geneva Planet Search program with \coralie{} spectrometer. \cite{McCarthy2004} determined the orbital period of the outermost planet \hd{}c $\simeq 3000$~days and large eccentricity $e_{\idm{c}} \sim 0.57$, however rendering the system unstable. The same year, \cite{Santos2004} detected  $\simeq 14$ Earth-mass planet \hd{}d in $\simeq 9.6$~d orbit with \harps{} spectrometer, achieving precision $\simeq 1$~m/s, actually below the RV variability (aka stellar jitter) induced by the Sun-like stars themselves. Furthermore, \cite{Butler2006} published 108 new observations of \muarae{}, spanning about of 7.5~yr,  made after AAT \ucles{} update, also approaching the measurement uncertainty below $1$~\ms{} at the end of the observational window. Shortly, \cite{Pepe2007} published RVs from their \harps{} followup, and announced the discovery of the fourth, Saturn-mass planet in the system. In parallel, \cite{Gozdziewski2007a} independently used genetic algorithms to re-analyse data in the \cite{Butler2006} catalogue, and they found a very similar solution with small eccentricity orbits, also including the fourth planet with the orbital period $\simeq 307$~days. That planet ``hided'' in the RV signal, because this period is approximately two times shorter as {that of the firstly detected} planet~\hd{}b.  Such a planet was unexpected in the paradigm of characterizing planets in order correlated with their RV variability. \cite{Gozdziewski2007a} concluded that the four-planet system may be long term stable in a wide range of the outermost period. However, it could not be constrained very well at that time, in $\simeq 3000$--$5000$~days range. 

Since then, the star has continued to be RV-monitored. The \harps{} measurements are now publicly available in the RV catalogue from archival spectra carefully reduced by  \cite{Trifonov2020}. Also, very recently \cite{Benedict2022} published additional 180~measurements from the \ucles{} spectrometer. The data altogether span 17.3~years ($\simeq 6318$~days), between epochs \epk{} and \epl{}. \cite{Benedict2022} {aimed to derive the new solution for the system based on} combined RVs with Hubble Space Telescope (\hst{}) astrometry. {They investigated possible astrometric signals of the planets.   They conclude that the residuals $\simeq 1$-$2$~mas to the canonical 5-parameter astrometric model contain marginal or no evidence for any of the planets in the \muarae{} system, making it possible only to constrain lower masses of the planets to $4$-$7\,\mJ{}$ (i.e., 2-3 times larger than the minimal masses estimated with the RVs).
} 

Furthermore, \cite{Benedict2022} report their updated {Keplerian} RV solution including the Saturn-mass planet as catastrophically unstable. They conclude that a notorious instability problem of the system remains unsolved{, invoking  \cite{Pepe2007,Laskar2017,Agnew2018} and \cite{Timpe2013}. }  This renewed our interest in the dynamics of \muarae{} system, given {simultaneously our} earlier, extensive investigations \citep{Gozdziewski2003e,Gozdziewski2005}, and the results in \citep{Gozdziewski2007a}. {We found quite an opposite conclusion} that the four-planet architecture, and moderate eccentricity of all planets is {\em crucial} to maintain the long-term stability of the system. {Actually, we found in \citep{Gozdziewski2007a} that the 3-planet model involving only two outer Jovian planets is localised at the very border of dynamical stability, with planets in high-eccentricity orbits, and such a feature indicated that the adopted model was incomplete or incorrect.} 

Extending the RV time series puts the long-term monitored planetary systems deeper in the stability zone. A recent discussion of this heuristic effect can be found in \citep{Stalport2022}. What is more, not only the RV data covers twice the time range in earlier work. The most accurate \harps{} data recently been independently reprocessed using a new RV pipeline by \citep{Trifonov2020}. They discovered and removed various systematic errors in a large sample of spectra. In some cases, they claim, the new RVs with improved accuracy can lead to orbital solutions different or more accurate from those found so far, including the hope of detecting additional planets. All of this gives us ample opportunity to test earlier predictions. Our goal is also to update the system's position in stability diagrams and statistics of multiple systems, studied for example by \cite{Timpe2013} and \cite{Laskar2017}. 

In addition to explaining this {\em qualitative} discrepancy between the results in \citep{Benedict2022} and in \citep{Gozdziewski2007a} the motivation for this work is to answer several open questions which have not been previously addressed in the literature.
 
Since that the current RV data covers almost twice the observational window since 2006, we want to constrain the orbit of Jupiter's outermost planet. It was determined with a large uncertainty of 700~days reported in \citep{Pepe2007} and an even larger uncertainty of $\pm 1300$~days in \citep{Gozdziewski2007a}.

Also, it is known that a sufficiently long interval of RVs data makes it possible to detect gravitational interactions between the planets \citep[e.g.][]{Laughlin2001}. Until now, the RVs of \muArae{} have been modeled in terms of a Keplerian parameterization of the orbital elements, since the interactions of its planets were not measurable at the time. In this kinematic approach, the inclination of the system remains completely unbounded. However, the most accurate Newtonian model can break the mass-inclination degeneracy, or at least constrain the masses of the planets indirectly through the stability requirement.

Our goal is also to resolve the open question of whether the inner Saturn-Jupiter planet pair is involved in the 2e:1b~MMR, or whether it is only close to this resonance. As far as this is concerned, the conclusions in both \citep{Pepe2007} and \citep{Gozdziewski2007a} were uncertain, as both types (resonance or near-resonance) of solutions were possible. However, this is crucial for explaining the apparent excess of planet pairs near low-order resonances \citep[e.g.,][and references therein]{Petrovich2013,Marzari2018}. The detailed characterization of multiple planetary systems, including their orbital resonances, is one of the fundamental problems from the point of view of the theory of planet formation and for explaining their observed orbital architectures. 

If our early predictions in \citep{Gozdziewski2007a} hold, and we find a dynamically stable orbital architecture for the planets, it may be possible to study the structure of debris disks in the system, particularly in the broad zone between 1.5~au and 5.2~au, and beyond the outermost planet. According to the packed planetary systems (PPS) hypothesis \citep[][and references therein]{Barnes2007}, smaller planets may exist in the system, but below the current RV detection level, approximately $1$\,\ms{}, which correspond to the Earth's mass range.

Finally, the highly hierarchical configuration of the \muarae{} planets imposes numerical problems in studying the long-term stability of the system, either through direct numerical integrations or by using the fast indicator approach, which is preferred in this work. Recall that the system contains a warm Neptune in an orbit of 9.6~days, as well as a very distant companion in an orbit of $4100$~days, forcing a huge reduction in the discretization step size.  To solve this problem, we propose a new numerical algorithm called \rem{} \citep{Panichi2017}, which we proved to be a close analogue of the Maximum Lyapunov Exponent (\mle{}). In this work, we compare the results of this fast indicator with the well-tested and widespread \megno{} \citep{Cincotta2003,Gozdziewski2001a}. We show that despite simplicity of the algorithm, the \rem{} indicator yields 1:1 dynamic maps compared to \megno{} and still outperforms the later variational algorithm in terms of CPU overhead.

We attempt to answer the questions posed above from the perspective of both updated RV time series and constraints provided with astrometric observations, as well as new statistical formulations of the RV model, dynamic and computational tools that have emerged over the time since the studies of \cite{Gozdziewski2007a} and \cite{Pepe2007}; we note that \cite{Benedict2022} also modeled the RV using the former, now somewhat ``outdated'' approach.

Planets discovered in the \muArae{} system are named in different ways. Here we adopt three designations: the first one is based on the star name, as the central object and subsequent Roman letters (``b'', ``c'', ``d'', and so on) attributed to the planetary companions in the chronological order of their discovery \citep{Gozdziewski2007a}. The second method is to enlist the planets according to their distance from the star, with digits ``1'', ``2'', ``3'', and so on. Finally, we use the names attributed to the planets by the International Astronomical Union (2015) in the NameExoWorld campaign\footnote{\url{https://www.nameexoworlds.iau.org/}}, among firstly discovered 19 extrasolar planetary systems. They were inspired by characters from the famous {\em Don Quixote} book by Miguel de Cervantes. So the \muArae{} system is composed of the host star Cervantes (\hd), and planetary companions \hd{}d (\araeD, planet~``1''), \hd{}e (\araeE, planet~``2''), \hd{}b (\araeB, planet~``3''), and \hd{}c (\araeC, planet~``4''), respectively.

The paper is structured as follows. After this Introduction, we describe data sources used for this study in Sect.~\ref{sec:data}. We {discuss planet detection limits, based on the astrometric \hst{} data and their analysis reported in \citep{Benedict2022}, as well as our independent simulations of the astrometric signal.} In Sect.~\ref{sec:models} we briefly recall essential details on the RV modeling in terms of Keplerian and Newtonian parameterization of the initial conditions (\ics{}) for multi-planet configurations, {and we point out factors omitted in the prior literature}. We report on a comparison of the results based on these two RVs parametrizations. {Sect.~\ref{sec:stability} is devoted to the long-term stability of the system. We aim to bound the inclination of the system with the RVs alone, based on the Newtonian model and statistical and dynamical constrains}. Section~\ref{sec:debris} is devoted to numerical simulations that reveal the dynamical structure of hypothetical debris disks in the system {as well as indicate possible localization of additional smaller planets}. {The work is summarised in Conclusions.} 

%_______________________________________________________________________________
%
\section{The reflex motion data for \muarae{}}
\label{sec:data}
%_______________________________________________________________________________
%_______________________________________________________________________________
%
{\subsection{Astrometric observations}}
%_______________________________________________________________________________
%

\begin{figure*}
\centerline{
\hbox{
\includegraphics[width=0.48\textwidth]{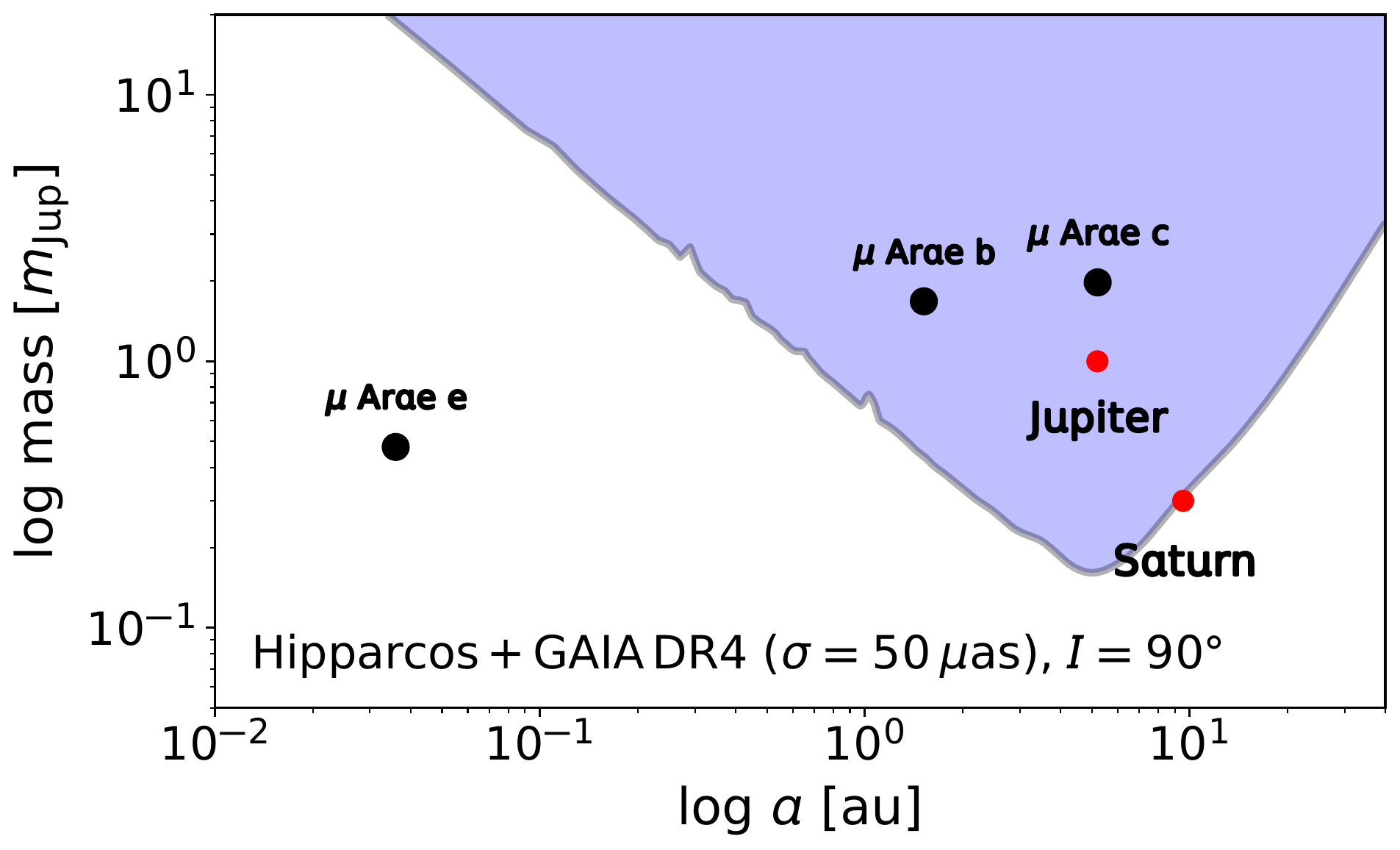}%{../disks/DiskXY.png}
\quad
\includegraphics[width=0.48\textwidth]{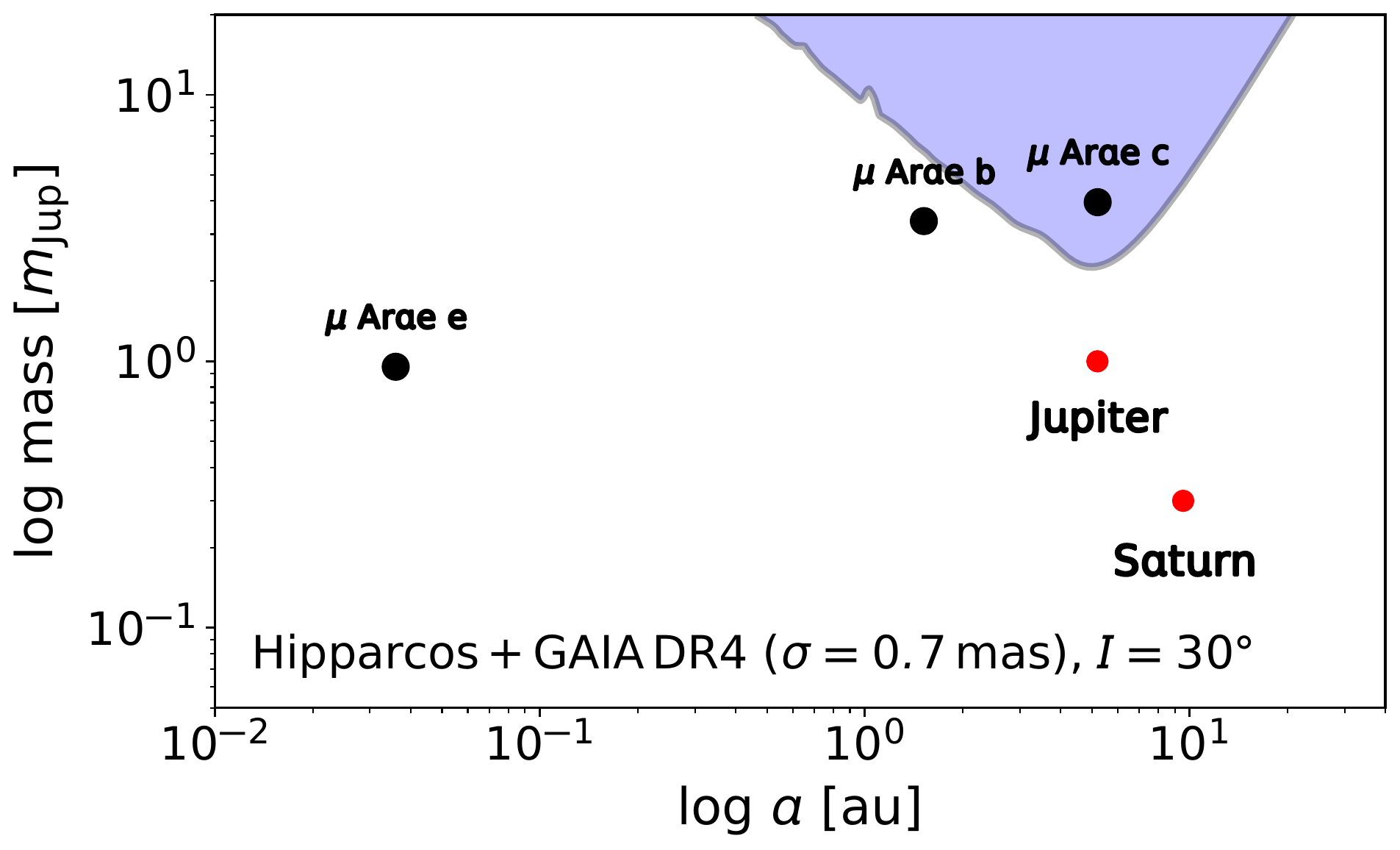}%{../disks/DiskXY.png}
}
}
\caption{
Astrometric detection limits in the mass--semi-major axis space for planets in circular, edge-on orbits simulated with the \htof{} package \citep{Brandt2021}, based on perturbed motion of the star due to the presence of planets. Objects in the blue--shaded region would be detected within the $\Delta\chi^2>30$ criterion when combining \gaia{}\,DR4 and \hipparcos{} \iad{} . (This criterion assumes $\Delta\chi^2=0$ for a free, inertial motion of the star). The left panel is for edge-on orbits of the Saturn- and Jupiter--like planets around \muArae{} with the orbital elements listed in Table~\ref{tab:tab2}, Fit~IIN.  Jupiter and Saturn are marked for a reference. We assume extremely high-precision \iad{} in the anticipated \gaia{}\,DR4 catalogue, with the mean uncertainty of $50\,\mu$as. The right panel is for the planet masses enlarged by the factor {$1/\sin(30^{\circ})$}, and artificial \iad{} accuracy of \gaia{}\,DR4 $\simeq 0.7$~mas, compatible with the declared \hst{} \fgs{} measurement precision reported in \citep{Benedict2022}. There are planned {96}~\gaia{} observations by the year of 2022, based on the \gaia{} Observation Forecast Tool (\gost{}).
}
\label{fig:fig1}
% figure 1
\end{figure*}

\cite{Benedict2022} observed \muarae{} with the \hst{} Fine Guidance Sensor (\fgs{}) between dates 2007.5 to 2010.4 (for about of 2~orbital periods of \muarae{}b). They made a detailed reduction of the observations and reported the results. Overall, the accuracy of the astrometric measurements $\simeq 0.6$--$0.7$~mas, and the residuals to 5-elements canonical astrometric solution (no companions present) are estimated on the level of $\simeq 1$--$2$~mas. However, the periodogram analysis of these residuals, which might contain unmodeled {factors} and a signature of companions, does not show any significant period overlapping with the known orbital variability from the RV analysis. Unfortunately, also analysis of the proper mean motion based on the \hst{} measurements by \cite{Benedict2022}, and \hipparcos{} \citep{vanLeeuwen2007} by \cite{Brandt2021c}, respectively, relative to the estimates in the \gaia{}\,DR3 catalogue indicate that there is {a marginal or lack of a measurable} difference between the proper mean motion at {the initial and the final} epochs for 25~years. That means there is {difficult to detect a significant} acceleration caused by the {planetary} companions, which was used, for instance, to astrometrically constrain the mass of the innermost planet~\hr8799{}e in \citep{Brandt2021a}. 

Given the negative detection of any of the companions, \cite{Benedict2022} estimated the lower mass limits for \muarae{}b,e,c as $(4.3, 7.0, 4.4)\,\mJ{}$, respectively, {which could be} consistent with a low inclination of the system below $I=30^{\circ}$. Moreover, they claim that inclinations in their sample of multiple-planetary systems are biased towards small values, $I \simeq 30^{\circ}$ and less. {As we show below, for \muarae{} this can be verified based of the RVs data alone.}

Although the parallax of the system is large, $\Pi\simeq 64$~mas, the relatively small semi-major axes of the planets, compared to other astrometrically detected systems, translate to weak astrometric signals. To illustrate this effect, and to predict if the system may be characterised astrometrically by the ongoing \gaia{} mission, we simulated detection limits with the Intermediate Astrometric Data (\iad{}) from the \hipparcos{} and \gaia{} surveys. For this purpose, we used the \htof{} package by \cite{Brandt2021} which makes it possible to combine data from both missions, including \iad{} for \gaia{} simulated with the help of \gaia{} Observation Forecast Tool \cite[][\gost{}]{Gaia2021}.

The results are illustrated in Fig.~\ref{fig:fig1}. The left panel is for the detection limits for outer, massive planets assuming that the inclination $I=90^{\circ}$ and masses are minimal (a less favorable scenario). Then, assuming a superior mean accuracy of $\simeq 96$~\gaia{} measurements scheduled by the end of 2022, with the mean uncertainty $\sigma \simeq 50$~mas in the anticipated DR4 catalogue, and \iad{}s from \hipparcos{}, we would easily detect the outermost pair of Jupiters. Note that the border of detection zone marks the astrometric detection criterion of $\Delta\chi^2>30$ by Perryman \citep{Brandt2021}, when $\Delta\chi^2=0$ applies to the free motion of the star. However, the inner Saturn-like planet remains deep below the detection limit (blue-shaded region). 

The situation is dramatically worse, if a hypothetical data accuracy $\simeq 0.7$~mas is close to the \hst{} \fgs{} astrometry. Even if the system inclination is statistically most likely for $I=60^{\circ}$ {or smaller, consistent with the inclination bias reported in \citep{Benedict2022}, $I=30^{\circ}$, scaling the minimal masses by a factor of $\simeq 20\%$ and $\simeq 100\%$, respectively, only the outermost planet could be barely detected with the astrometric time-series.}

Unfortunately, these arguments and simulations leave little hope that a re-analysis of the available astrometric data may change the results and conclusions in \citep{Benedict2022} and \citep{Brandt2021c}. Therefore we abandoned the \hst{} astrometry from further analysis, and we focused on the RV observations only.

%_______________________________________________________________________________
%
\subsection{Radial Velocity data}
%_______________________________________________________________________________
%
We considered two slightly different sets of the RV measurements for \muArae{} available in public archives and sources.

The RV data set~\Done{} consists of 380 measurements spanning 6317.5~days. They are collected with three instruments: \coralie{} (${\cal D}_{\rm\coralie{}}$), \ucles{} (${\cal D}_{\rm\ucles{}}$) and \harps{} (${\cal D}_{\rm\harps{}1,2}$). This set is literally the same as in \cite{Benedict2022}, and we obtained it from the author (private communication). In densely sampled parts of the observational window, the data were binned if there was more than one measurement made during a night. The mean uncertainty is different for individual spectrometers, and varies between  $\mean{\sigma} \sim 1$~\ms{} up to a several \ms{} for \coralie{}. Moreover, \cite{Benedict2022} considered \harps{} observations in two disjoint sets: from \cite{Pepe2007} and the second part of the time-series after that date from \citep{Trifonov2020}. They attributed different RV offsets to these sets. 

We also compiled a second data set~\Dtwo{}. \cite{Trifonov2020} derived the RV velocities from spectra obtained prior-- and post-- the \harps{} upgrade in May 2015, and corrected them for various systematics and instrumental effects. Since the available data for \muarae{} contains effectively only two post-upgrade measurements made in nights of June and July 2015, we skipped these points from the orbital analysis. It would be difficult to account for two free parameters, $\sigma_f$ and $V_0 \equiv V_{0,{\rm\ucles{}}}$, to be statistically determined with the RV subset comprising of only two datum. Moreover, because the post-upgrade \harps{} epochs overlap with \ucles{} measurements, skipping them unlikely may change the model results. We also get rid of two free parameters. Similarly to \cite{Benedict2022}, we also binned densely sampled measurements, but with a smaller interval of 0.1~days. Before doing that, we removed several points from the \harps{} RV time series in \citep{Trifonov2020}, with heavily outlying uncertainties of $10$--$24$\,\ms{}, given the mean uncertainty $\sigma_{\rm\harps{}} \simeq 1$\,\ms{}. The problematic measurements appear around JD\,2453169 (mid-June, 2004), when literally hundreds of spectra were taken overnight. Removing these points should not cause any problem, due to the dense sampling and binning. For the binned data in set~\Dtwo{}, we adopted the uncertainties as the mean uncertainty in a particular bin.

In this way, the data set \Dtwo{} consists of the whole pre-upgrade \harps{} measurements ${\cal D}_{\rm\harps{}}$, as a homogeneous data set from \cite{Trifonov2020}, and ${\cal D}_{\rm\coralie{}}$ and ${\cal D}_{\rm\ucles{}}$ from \cite{Benedict2022}. This set has 411~measurements and also spans 6317.5~days. To simplify presentation of the RV offsets, we subtracted the mean value of all RVs in a given subset from individual RVs in this subset.

{Finally, in some experiments we considered data set \Dthree{} composed of 349 measurements from the pre-upgrade \harps{}  and ${\cal D}_{\rm\ucles{}}$ from \cite{Benedict2022}. These RV time-series span
the same time interval as \Dtwo{} does. This data set \Dthree{} lacks the less accurate ${\cal D}_{\rm\coralie{}}$ RVs.
}
%_______________________________________________________________________________
%
%_______________________________________________________________________________
%
\subsection{Keplerian vs Newtonian Radial Velocities}
%_______________________________________________________________________________
%
\label{sec:models}
The mathematical models for the RV velocities are well known. However, to keep the presentation self-consistent, and to cover some nuances, we will briefly recall the required material.

Since, following the prior literature,  we expect that the \muArae{} orbits may be quasi-circular, to get rid of weakly constrained longitudes of pericenter $\varpi_i$ when eccentricities $e_i\sim 0$, we  introduce {Poincar\'e{} elements} 
$\left\{x_i = e_i \, \cos\varpi_i, y_i = e_i \, \sin\varpi_i \right\}$, $i=1,2,3,4$. 
Also, the mean anomaly ${\cal M}$ at the initial epoch $\tepoch{}$ denoted as $\Mmean_i \equiv \Mmean_i(t_0)$ is defined through the III law of Kepler, but written for the Jacobian reference frame
\begin{equation}
\label{eq:kepler}
P_i = 2\,\pi \sqrt{\frac{a_i^3}{k^2\,(m_0 + m_1+ \ldots m_i)}}, \quad
{\cal M}_i(t) = {\cal M}_i +  \frac{2\pi}{P_i} \left( t-t_0 \right),
\end{equation}
where $k$ is the Gauss constant, and $P_i$, $a_i$ stand for the orbital period and semi-major-axis for each planet, respectively. 

Regarding the Keplerian parameterization of the RV, we apply the well known canonical formulae \citep{Smart1949}  {due to the presence of planets}
\begin{eqnarray}
\label{eq:kep}
 \mbox{C}(t) &\equiv& V_r^{\rm K}(t) = 
 \sum_{i}^{\Npl} K_{i} \left[ e_i \cos\omega_i + \cos(\nu_i +\omega_i) \right], \\
 &=& \sum_{i}^{\Npl} K_{i} [ x_i + (x_i^2+y_i^2)^{-1/2} \left( x_i \cos \nu_i - y_i \sin \nu_i \right) ],  
\end{eqnarray}
{where $\omega \equiv \varpi$ for a coplanar system},  $\nu \equiv \nu(t)$ denotes the true anomaly of a~planet, $\Npl$ is the number of planets in the system, and $\nu=\nu(P,e,{\cal M}(t))$. To characterize the orbit of the $i$-th planet, we need to know five free orbital elements:  $\vec{\theta}_i = \left[ K_i, P_i, x_i \equiv e_i\,\cos\varpi_i, y_i \equiv e_i \, \sin\varpi_i, {\cal M}_i \right]$, where the RV semi-amplitude $K_i$ depends on the minimal mass of the planet $m_i \sin I$, when the inclination $I=90^{\circ}$.

Let us note that we interpret the RV signal in terms of the geometric elements inferred in the Jacobian frame of reference. We follow here conclusions and discussion in \cite{Lee2003}, to properly express parameters of the Keplerian model through the $N$-body initial condition. We need that to investigate the long-term stability of the system with the numerical integrations.  For relatively massive planets, the Jacobian (canonical) elements account for indirect interactions between the planets on Keplerian orbits to the first order in the masses (the ratio of planet masses to the star mass),  see also \citep{Gozdziewski2012} for more details. 

In order to derive the $N$-body initial condition from the fitted Keplerian elements $\vec{\theta}_i$, $i=1,\ldots,\Npl{}$, we first determine the minimal masses $m_i \sin I \equiv m_i$  and semi-major axes $a_i$ of the planets. The semi-amplitude $K_i$ of the RV signal
\[
 K_i \sqrt{1-e_i^2} = a_i \left( \frac{2\pi}{P_i} \right)  
 \frac{m_i}{(m_0 + m_1 + \ldots + m_i)},
\] 
where the $a_i$ constrained by the observationally derived orbital period $P_i$ obeys Eq.~\ref{eq:kepler}, and $m_0$ stands for the star mass.
%
%\[
% \left(\frac{2\pi}{P_i}\right)^2 a_i^3 = k^2 (m_0 + \ldots + m_i), 
%\]
%and $k$ is the Gaussian gravity constant. 
%
Eliminating $a_i$, we obtain a {cubic equation for the unknown masses}, which may be subsequently solved for $m_i$, $i=1,2,\ldots$, {based on analytical formulae or with a simple Newton-Raphson scheme (a few iterations suffice to reach the machine accuracy)}. Then we transform the geometric elements to Cartesian coordinates and velocities with the standard two-body formulae, where the gravitational parameter for the $i$th planet is $\mu_i = k^2 ( m_0 + m_1 + \ldots m_i)$. 

To determine parameters of the orbital model explaining the RV time-series, we optimized a canonical form of the maximum likelihood function ${\cal L}$ \citep{Baluev2009}:
\begin{equation}
 \ln {\cal L} =  
-\frac{1}{2} \sum_{i,t}
\frac{{\mbox{(O-C)}}_{i,t}^2}{\sigma_{i,t}^2}
- \frac{1}{2}\sum_{i,t} \ln {\sigma_{i,t}^2} 
- \frac{1}{2} \Nm{} \ln{2\pi},
\label{eq:Lfun}
\end{equation}
where $(\mbox{O}-\mbox{C})_{i,t}$ is the (O-C) deviation of the observed $t$-th RV observation, with the uncertainty $\sigma_{i,t}^2 \rightarrow \sigma_{i,t}^2+\sigma_f^2$, with $\sigma_f$ parameter scaling the raw error $\sigma_{i,t}$ in quadrature, and $\Nm{}$ is the total number of the RV observations. We assume that the uncertainties are Gaussian. 

The error floor factors $\sigma_f^2$ are different for each telescope, as they may involve not only the intrinsic, chromospheric RV stellar variability (stellar jitter), but also an instrumental uncertainties inherent to each telescope and the RV pipeline. The RV model also involves individual offsets of the zero-level RV for each instrument. Distinguishing between these two parameters is important even for the same spectrometer and different setups of its work. For instance, the upgrade of \harps{} optical fibres around the middle of~2015 changed the instrumental profile and thus the RV offset between the pre- and post-upgrade RVs. To complicate things even more, the RV offset may be not the same for all stars and may even depend on the stellar spectral type \citep{Trifonov2020}. 

{Therefore fitting the jitter uncertainties as free parameters of the model is crucial to obtain adequate statistical representation of the RV data. We may note here, that in the past, these parameters have been fixed based on the averaged values for chromospherically quiet stars of a given spectral type. That recently outdated (and somewhat incorrect) approach was used by \cite{Gozdziewski2007a} and \cite{Pepe2007}; \cite{Benedict2022} tuned {the RV uncertainties to obtain} $\Chinu \simeq 1$.}

Usually, the Keplerian model determines sufficiently accurately the $N$-body, exact RVs. However, for systems with large-mass planets, this equivalence may be questionable, especially if the interval of the RV time series becomes long. Then we have to introduce the self-consistent model that requires solving the Newtonian equations of motion. The RV {due to the planets} is the velocity component of the star along the $z$-axis w.r.t. the barycenter of the Solar system
\begin{equation}
\label{eq:newt}
  \mbox{C}(t) \equiv V_r^{\idm{N}}(t) =
   -\frac{1}{m_0} \sum_{i=1}^{\Npl} {m_i} {\dot z}_i(t),
\end{equation}
which is parameterised through planet masses and the osculating orbital elements 
$\vec{\theta}_i = \left[ m_i, a_i, x_i, y_i, {\cal M}_{i} \right]$ for each planet in the system. Here, as the osculating epoch we select the epoch of the first observation in the given time series. {In some experiments, we also selected the osculating epoch in the middle of the data window.} 

Expressions for the RVs, Eq.~\ref{eq:kep} and Eq.~\ref{eq:newt} have to be accompanied with the instrumental zero-level offset $V_{0,j}$, $j=1,\ldots,M$ that makes it possible to compute $(\mbox{O}-\mbox{C})(t)$ in Eq.~\ref{eq:Lfun}. 
{
For $\Npl$-planets forming a coplanar system observed with $M$ instruments, we have therefore $p = 5 N + 2 M$ free parameters to be fitted to one-dimensional time series of the RV observations.
}

The definition in Eq.~\ref{eq:Lfun} is constructed so the best-fitting models should yield  $\Chi = \chi^2/(\Nm{}-p) \sim 1$, and $\chi^2_{\nu}$ cannot be used to compare the models quality.  Instead, \cite{Baluev2009} proposed to use: 
\[
 \ln L  = -\ln{\cal L}/\Nm{} - \ln(2 e\pi)/2,
\]
where $L$ is expressed in ms$^{-1}$. This statistics is suitable to assess the relative quality of fits, since $L \sim \mean{\sigma}$ measures a scatter of measurements around the best-fitting models, similar to the common RMS {--- smaller $L$ means better fit}. 

In order to localize the best-fitting solutions in the multi-dimensional parameter space, we explore it with evolutionary algorithms \citep[GEA from hereafter,][]{Charbonneau1995,Izzo2010}.  We then perform the MCMC analysis in the neighborhood of selected solutions using an affine invariant ensemble sampler \citep{Goodman2010} encompassed in a great \emcee{} package \citep{Foreman2014}.  The computations were performed in multi-CPU environment, making it possible to evaluate 128,000--256,000 (or more) of 144--384 \emcee{} ``walkers'' from a small-radius ball around a solution found with the GEA.
%
%Recalling briefly, the Markov Chain Monte Carlo (MCMC) technique is widely used by the astronomic community to determine the posterior probability distribution ${\cal P}(\tv{\theta}|{\cal D})$ of model parameters $\tv{\theta}$, given the data set ${\cal D}$: 
%$
%  {\cal P}(\tv{\theta}|{\cal D}) 
%  \propto {\cal P}(\tv{\theta}) {\cal P}({\cal D}|\tv{\theta}),
%$
%where ${\cal P}(\tv{\theta})$ is the prior, and the sampling data distribution
%${\cal P}({\cal D}|\tv{\theta}) \equiv \log{\cal L}(\tv{\theta},{\cal D})$.

We select all priors as flat (or uniform, improper) by  sufficiently broad ranges on the model parameters, e.g., $P_{i} \in [1,10,000]$~days, $x_i,y_i \in [-0.25,0.25]$, $m_i \in [0.1, 14]~\mJ$, ($i=1,2,3,4$), the error floors (jitters) $\sigma_{f,j}>0$~\ms{}, $j=1,\dots, M$. In a few experiments with the $N$-body model, we also tested Gaussian priors for the $({x_1,y_1})$ elements of the innermost planet, with the mean equal to zero and variances $\sigma_{x,y} = 0.05, 0.075, 0.1$, respectively.  In this case, however, the results of sampling did not substantially change, compared to the flat priors.

\subsection{The best-fitting orbital configurations}
%_______________________________________________________________________________
%
We first performed an extensive search for the best-fit solutions using GEA, and we collected $\simeq 10^3$ solutions for both data sets and model variants. We found that the best-fit Keplerian and Newtonian models with $L \simeq 3.2$~\ms{} (RMS $\simeq 3.4$~\ms{}) have well determined extrema of $\ln {\cal L}$ for orbital periods $P_i$ of roughly $9.64, 308, 645, 4030$~days, respectively. Also, all osculating eccentricities are well limited to moderate values, roughly in the range of $0.02$--$0.1$.

\begin{table*}
\caption{
Best-fit parameters of the $\mu$~Arae (Cervantes) system for the Keplerian (Fit~IK) and Newtonian (Fits~IN) parameterization, data set \Done{}. The osculating epoch is the date of the first observation in the \ucles{} data set. The system is coplanar with the inclination $I=90^{\circ}$ and nodal longitudes $\Omega=0^{\circ}$. The stellar mass is $1.13\,\msun$ \citep{Bonfanti2015} as used in \citep{Benedict2022}, close to $1.10\pm 0.02\,\msun$ in \citep{Soriano2010}. The RV offsets are computed w.r.t. the mean RV in each individual data set. Uncertainties are estimated around the median values $\mu$, i.e., $[\mu-\sigma,\mu+\sigma]$ as the 16th and 86th percentile of the samples. Numerical values forFit~IN selected from the MCMC samples with low RMS are quoted to the 7th digit after the dot, to make it possible to reproduce the dynamical maps and direct numerical integrations. The mean longitude $\lambda=\varpi+{\cal M}$ at the epoch was computed from the MCMC samples.
}
\label{tab:tab1}
\begin{tabular}{l r r r r}
\hline
Planet           
& \muarae{}d (\araeD, 1)   & \muarae{}e (\araeE, 2) & \muarae{}b (\araeB, 3) & \muarae{}c (\araeC, 4) \\
\hline\hline
\multicolumn{5}{c}{Fit~IK (Keplerian model of the RV, data set \Done{}, RMS $=3.4$\ms{})} \\
\hline
$K$\,[\ms{}]     &  2.95$\pm$0.19   &  13.22$\pm$0.34  & 36.47$\pm$0.22  & 23.17$\pm$ 0.33  \\
$P\,$[d]         &  9.638$\pm$0.001 & 308.75$\pm$0.29  & 645.00$\pm$0.36 &   4060$\pm$ 27    \\
$e \cos\varpi$   & -0.104$\pm$0.063 & -0.093$\pm$0.014 & 0.058$\pm$0.011 & 0.022$\pm$ 0.012 \\
$e \sin\varpi$   & -0.059$\pm$0.063 & -0.014$\pm$0.017 & 0.023$\pm$0.008 & 0.032$\pm$ 0.013 \\
$e$              &  0.137$\pm$0.056 &  0.096$\pm$0.014 & 0.063$\pm$0.010 & 0.040$\pm$ 0.013  \\
$\varpi\,$[deg]  & 210$\pm$32  &  189$\pm 10$  & 21.6$\pm$(9.4,8.2)  & 55.9 $\pm$17.5 \\
${\cal M}$\,[deg]& 223.3$\pm$32 & 66.7$\pm$10.5 & 272.5$\pm$(8.3,9.4) & 185.9$\pm$ 17.2 \\
$\lambda$\,[deg]& 73.3$\pm$10.5 & 255.4$\pm$3.9 & 294.0$\pm$1.0 & 241.9$\pm$ 2.2 \\
$V_0$\,[\ms{}]  & \multicolumn{4}{c}{
                         \coralie{}: 13.04$\pm$0.42, \quad 
                          \ucles{}: -7.80$\pm$1.20, \quad
                          \harps{}$_1$:  1.0$\pm$0.3, \quad
                           \harps{}$_2$: -4.20$\pm$0.32
                }\\
 $\sigma_f$\,[\ms{}] & \multicolumn{4}{c}{
                        \coralie{}:   1.30$\pm$0.21, \quad 
                         \ucles{} :  6.1$\pm$1.1, \quad
                          \harps{}$_1$:  0.62$\pm$0.46, \quad
                          \harps{}$_2$ :  1.67$\pm$0.40
                }\\
\hline
\multicolumn{5}{c}{Fit~IN (Newtonian model of the RV, data set \Done{},  RMS $=3.4$\ms{})} \\
\hline
$m \sin I$\,[$\mJ$]   
                & 0.033$\pm$0.002 & 0.477$\pm$0.012 & 1.680$\pm$0.010 & 1.978$\pm$0.028\\
                & 0.0333733 & 0.4805150 &  1.6894371  & 1.9415698     \\
$a$\,[au]       & 0.092319$\pm$6$\times{}10^{-6}$ 
                                 & 0.9376$\pm$0.0015& 1.521$\pm$0.001 & 5.243$\pm$0.023\\
                & 0.0923201 & 0.9358533 &   1.5204938 & 5.2228363  \\              
$e \cos\varpi$  &-0.086$\pm$0.067 &-0.060$\pm$0.014 & 0.057$\pm$0.012 & 0.018$\pm$0.012\\
$e \sin\varpi$  &-0.063$\pm$0.067 &-0.031$\pm$0.015 & 0.016$\pm$0.008 & 0.026$\pm$0.012\\
$e$             & 0.127$\pm$0.057 & 0.069$\pm$0.014 & 0.060$\pm$0.011 & 0.034$\pm$0.012\\
                & 0.0093112 & 0.0729955    &  0.0563256 &   0.0378130\\
$\varpi\,$[deg] & 215$\pm$(36,38) & 207.5$\pm$(11.4,11.9) & 16.4$\pm$(10.3,8.2) & 56.4$\pm$21.0\\
                & 52.8721947   & 217.8362502     & 19.7788422     &  52.2928770 \\
${\cal M}\,$[deg]& 218$\pm$(34,38)  & 53$\pm$14 & 278$\pm$(9,10) &  187 $\pm$(21,20)\\
                & 25.8318188   & 36.6741123& 272.3695792& 187.6140820 \\ 
$\lambda\,$[deg]&76.9$\pm$10.7 &260.0$\pm$4.0 & 293.3$\pm$1.1 & 243.1$\pm$2.1\\                
$V_0$\,[\ms{}]   & \multicolumn{4}{c}{
                          \coralie{}: 13.10$\pm$0.43, \quad
                          \ucles{}:-7.74$\pm$1.14, \quad
                          \harps{}$_1$ 1.10$\pm$0.30, \quad
                          \harps{}$_2$: -3.94$\pm$0.32               
                }\\
$\sigma_f$\,[\ms{}]  & \multicolumn{4}{c}{
                         \coralie{}:    1.23$\pm$0.20, \quad   
                         \ucles{}:      5.88$\pm$(1.07,0.93) \quad
                         \harps{}$_1$:  0.45$\pm$0.40, \quad
                         \harps{}$_2$:  1.51$\pm$0.36
}\\
\hline         
\end{tabular}
\end{table*}

% Fit IN numerical values
% Mstar 1.13d0 msun  
%   m[mjup] a[au] e inc[deg]  om[deg]  Om[deg]  M[deg]  
% 0.0333733    0.0923201    0.0093112   0.0000000    0.0000000   52.8721947   25.8318188
% 0.4805150    0.9358533    0.0729955   0.0000000    0.0000000 -142.1637498   36.6741123
% 1.6894371    1.5204938    0.0563256   0.0000000    0.0000000   19.7788422  272.3695792
% 1.9415698    5.2228363    0.0378130   0.0000000    0.0000000   52.2928770  187.6140820
% 3.3733d-7     1.923201    0.0093112   0.0000000    0.0000000   -141.12579  206.74827
% -142.1637498 + 360 = 217.8362502000000
% -141.12579+360 ans =  218.8742100000000

\begin{table*}
\caption{
Best-fit parameters of the $\mu$~Arae (Cervantes) system for the Keplerian (Fits~IIK) and Newtonian (Fits~IIN) parameterization, data set \Dtwo{}. The osculating epoch is the date of the first observation in the \ucles{} data set. The system is coplanar with the inclination $I=90^{\circ}$ and nodal longitudes $\Omega=0^{\circ}$. The stellar mass is $1.13\,\msun$ \citep{Bonfanti2015} as used by \citep{Benedict2022}, close to $1.10\pm 0.02\,\msun$ in \citep{Soriano2010}. The RV offsets are computed w.r.t. the mean RV in each individual data set. Uncertainties are estimated around the median values $\mu$, i.e., $[\mu-\sigma,\mu+\sigma]$ as the 16th and 86th percentile of the samples. Numerical values for Fit~IIN selected from MCMC samples with low RMS are quoted to the 7th digit after the dot, to make it possible to reproduce the dynamical maps and direct numerical integrations. {The mean longitude $\lambda=\varpi+{\cal M}$ at the epoch was computed from the MCMC samples.}
}
\label{tab:tab2}
\begin{tabular}{l r r r r}
\hline
Planet           
& \muarae{}d (\araeD, 1)   & \muarae{}e (\araeE, 2) & \muarae{}b (\araeB, 3) & \muarae{}c (\araeC, 4) \\
\hline\hline
\multicolumn{5}{c}{Fit~IIK (Keplerian model of the RV, data set \Dtwo{}, RMS $=3.4$\ms{})} \\
\hline
$K$\,[\ms{}]     & 2.84  $\pm$0.17 &  12.36$\pm$0.30  & 35.81$\pm$0.20   & 22.7$\pm$ 0.26  \\
$P\,$[d]         & 9.638$\pm$0.001 & 308.36$\pm$0.29    & 644.92$\pm$0.29 &    4019$\pm$ 24    \\
$e \cos\varpi$   &-0.052$\pm$0.037 &  -0.073$\pm$0.014& 0.036$\pm$0.011  &-0.001$\pm$ 0.011 \\
$e \sin\varpi$   &-0.024$\pm$0.040 &  -0.012$\pm$0.017& 0.025$\pm$0.008  & 0.054$\pm$ 0.011 \\
$e$              & 0.071$\pm$0.034 & 0.076$\pm$0.014  & 0.045$\pm$0.008  & 0.055$\pm$0.011  \\
$\varpi\,$[deg]  & 204  $\pm$41&  189$\pm$13     &    35.1$\pm$(14.3,12.6)  &  91 $\pm$12 \\
${\cal M}$\,[deg]& 225 $\pm$(41,44) &  62$\pm$13  & 258.3$\pm$(12.5,14.3)  & 147$\pm$11    \\
${\lambda}$\,[deg]& 69.0 $\pm$10.5&  250.7$\pm$4.1  &   293.4$\pm$0.9  &  237.7$\pm$1.9    \\
$V_0$\,[\ms{}]  & \multicolumn{4}{c}{
                          \coralie{}: -7.36$\pm$1.10, \quad
                          \ucles{}:  0.77$\pm$0.25, \quad
                         \harps{}: 2.12$\pm$0.20
                          %\harps{}$_1$: -7.8$\pm$1.2, \quad                    
                }\\
$\sigma_f$\,[\ms{}]  & \multicolumn{4}{c}{
\coralie{}:  5.33$\pm$0.99, \quad    \ucles{}:  0.68$\pm$0.49, \quad
\harps{}1.80$\pm$0.14}\\
%RMS[\ms{}] & \multicolumn{4}{c}{3.4} \\      
%
%       
\hline
\multicolumn{5}{c}{Fit~IIN (Newtonian model of the RV, data set \Dtwo{},  RMS $=3.4$\ms{})} \\
\hline
$m \sin I$\,[$\mJ$]   
                 & 0.032$\pm$0.002 & 0.448$\pm$0.011 & 1.65$\pm$0.009  & 1.932$\pm$0.022\\
                 & 0.0297566 & 0.4558348 & 1.6608084 & 1.9478583 \\
$a$\,[au]        & 0.092319$\pm$5$\times{}10^{-6}$ 
                 & 0.9347$\pm$0.0015 & 1.522$\pm$0.001 & 5.204$\pm$0.021\\
                 & 0.0923174 & 0.9342193 & 1.5209196 & 5.2065203 \\       
$e \cos\varpi$   &-0.065$\pm$0.050 &-0.047$\pm$0.013 & 0.035$\pm$0.011 & -0.003$\pm$0.011 \\
$e \sin\varpi$   &-0.034$\pm$0.050 &-0.026$\pm$0.014 & 0.019$\pm$0.008 &  0.047$\pm$0.011 \\
$e$              & 0.090$\pm$0.042 & 0.055$\pm$0.014 & 0.041$\pm$0.009 &  0.049$\pm$0.011 \\
                 & 0.0172379 & 0.0447130 & 0.0423168 & 0.0242568 \\
$\varpi\,$[deg]  & 207$\pm$(39,41) & 209$\pm$(13,14)  & 28$\pm$(16,13) & 94.1$\pm$13.4\\
                 & 285.3319635 & 215.5470967 & 14.4134097 & 88.4886424 \\
${\cal M}\,$[deg]& 221$\pm$(40,44)  &  45$\pm$(16,15)   & 265$\pm$(14,16) &  145 $\pm$13\\
                 & 147.4681451 & 39.3680159 & 276.3668580 & 150.2821658 \\
${\lambda}\,$[deg]
      & 71$\pm$10  &  254.6$\pm$4.4  & 293.0$\pm$1.0 &  239.0 $\pm$1.9\\                 
$V_0$\,[\ms{}]   & \multicolumn{4}{c}{
                          \coralie{}: -7.2$\pm$1.1, \quad
                          \ucles{}:  0.87$\pm$0.26, \quad
                           \harps{}: 2.25$\pm$0.19                
                }\\
$\sigma_f$\,[\ms{}]  & \multicolumn{4}{c}{
\coralie{}:  5.3$\pm$1.0, \quad   
 \ucles{}:  0.48$\pm$0.42, \quad
                        \harps{}:   1.69$\pm$0.14
}\\
\hline
\end{tabular}
\end{table*}

%  rms  3.3169611899563280      
%  Mstar 1.13d0 msun + warm Neptune = 1.130028421d0   
%  m[mjup] a[au] e inc[deg]  om[deg]  Om[deg]  M[deg]
%    0.0297566    0.0923174    0.0172379   0.0000000    0.0000000  -74.6680365  147.4681451
%    0.4558348    0.9342193    0.0447130   0.0000000    0.0000000 -144.4529033   39.3680159
%    1.6608084    1.5209196    0.0423168   0.0000000    0.0000000   14.4134097  276.3668580
%    1.9478583    5.2065203    0.0242568   0.0000000    0.0000000   88.4886424  150.2821658
% octave:3> -74.6680365+360  = ans =  285.3319635000000
% octave:4> -144.4529033+360 = ans =  215.5470967000000

\begin{figure*}
\centerline{
\hbox{
\includegraphics[width=0.5\textwidth]{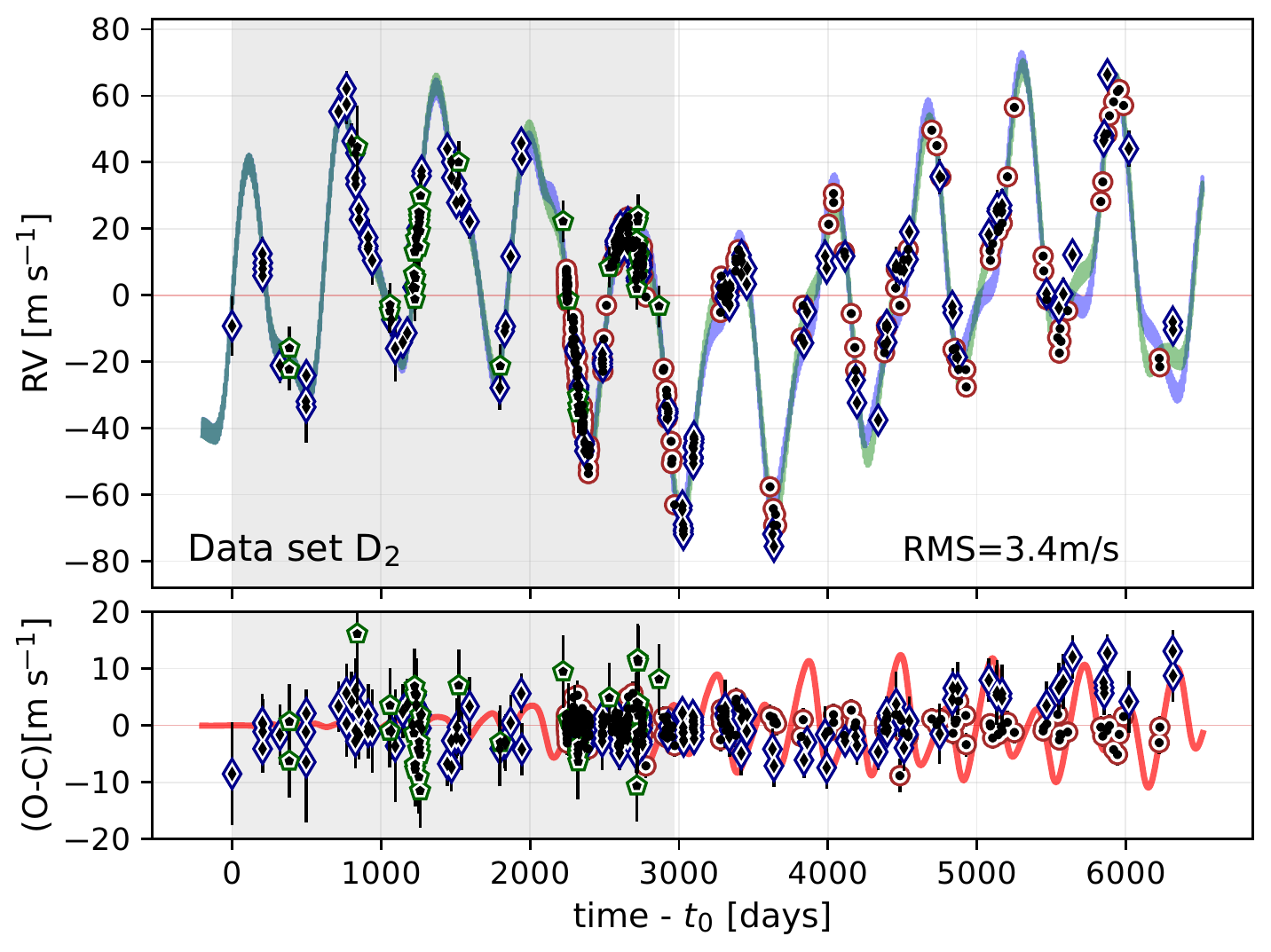} %{../Doppler/muAraeTest0.pdf}
\includegraphics[width=0.5\textwidth]{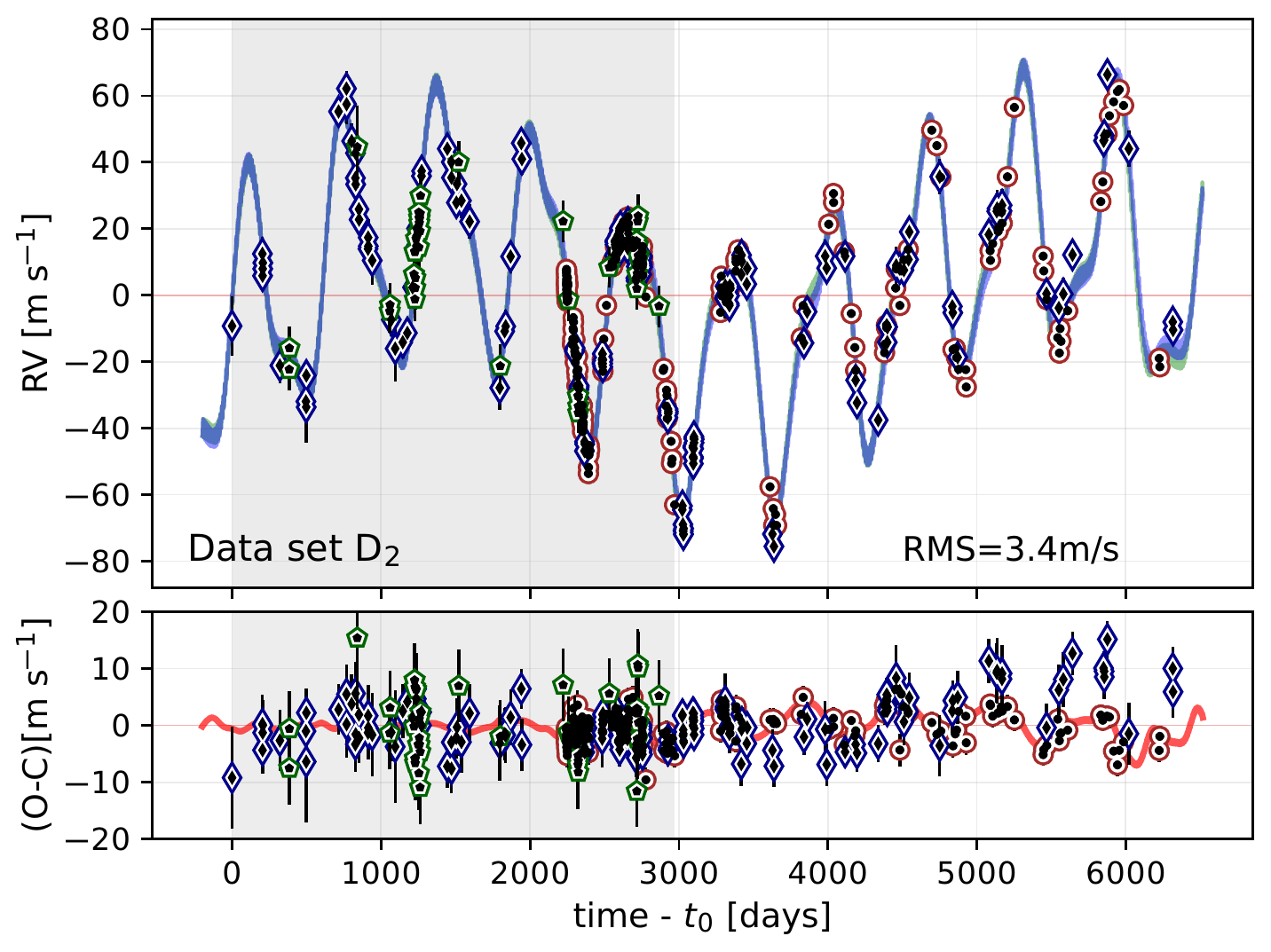} %{../Doppler/muAraeTest0.pdf}
}
}
\caption{
{\em Left panel}: synthetic curves of the best best-fitting Keplerian model to \Dtwo{} data set, depicted as Fit~IIK in Tab.~\ref{tab:tab2} (light-green curve) and its Newtonian interpretation (light-blue curve) over-plotted on the RV data. The difference between the RV signals illustrates a red curve in the Keplerian residuals (O-C) panel for the Keplerian ephemeris. Symbols describe the RV measurements from different spectrometers: green pentagons are for \coralie{}, brown/red circles are for \harps{} and blue diamonds are for \ucles{}. Error bars in the (O-C) diagram include the error floor parameters. The shaded rectangle marks the time-span of the RV data in \citep{Gozdziewski2007a} and \citep{Pepe2007}.
{\em Right panel}: 
Synthetic curves of best-fitting Keplerian (light-green curve) and Newtonian models (light-blue curve) to \Dtwo{} data set, depicted as Fit~IIK and Fit~IIN (Tab.~\ref{tab:tab2}), respectively, and over-plotted on the RV data. The difference between the signals illustrates a red curve in the Newtonian residuals (O-C) panel. Parameters of the models corresponds to the maxima of posterior samples. Symbols describe the RV measurements from different spectrometers: green pentagons are for \coralie{}, brown circles are for \harps{} and blue diamonds are for \ucles{}. Error bars in the (O-C) diagram include the error floor parameters, and the shaded region is the RV data time span prior to the analysis conducted in 2006.
}
\label{fig:fig2}
% figure 2
\end{figure*}

The resultig best-fitting parameters for data sets \Done{} and \Dtwo{} are given in Tables~\ref{tab:tab1} and~\ref{tab:tab2}. The best-fitting Keplerian model Fit~IIK in Tab.~\ref{tab:tab2} is illustrated in Fig.~\ref{fig:fig2}, left panel. Using this solution as an example, we checked the consistency of the Keplerian and Newtonian parameterization. We transformed Fit~IIK as osculating elements for the epoch of the first observation $t_0=$\epk{} in the \ucles{} data, as described in Sect.~\ref{sec:models}. We then computed the Newtonian RV signal through of numerical integration of the $N$-body equations of motion for the entire four-planet system with the IAS15 integrator \citep{Rein2015}. It turns out that the difference 
$
\Delta\mbox{RV}(t)=  V_r^{\idm{N}}(t) - V_r^{\idm{K}}(t)
$
increases in an oscillatory manner, reaching about $\pm 10$\ms{}, which exceeds more than twice the RV signal from the innermost planet (red curve in the residuals diagram in Fig.~\ref{fig:fig2}). 

\begin{figure*}
\centerline{
\vbox{
\hbox{
\includegraphics[width=0.5\textwidth]{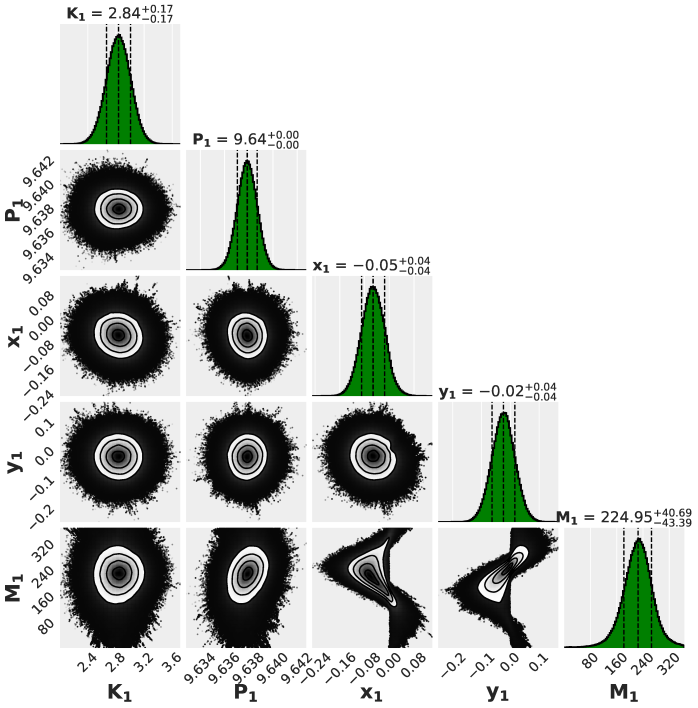}%{../Doppler7.Mcmc/muAraP1.pdf}
\includegraphics[width=0.5\textwidth]{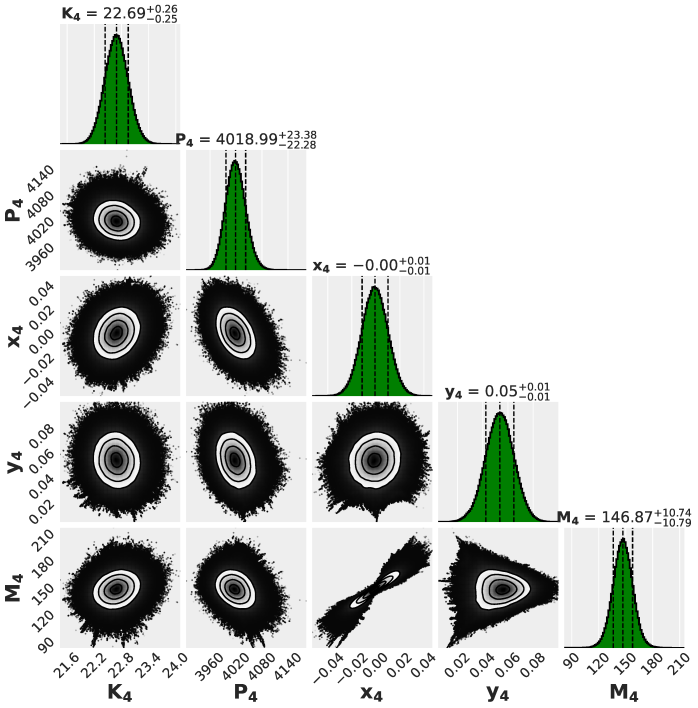}%{../Doppler7.Mcmc/muAraP4.pdf}
}
\hbox{
\includegraphics[width=0.5\textwidth]{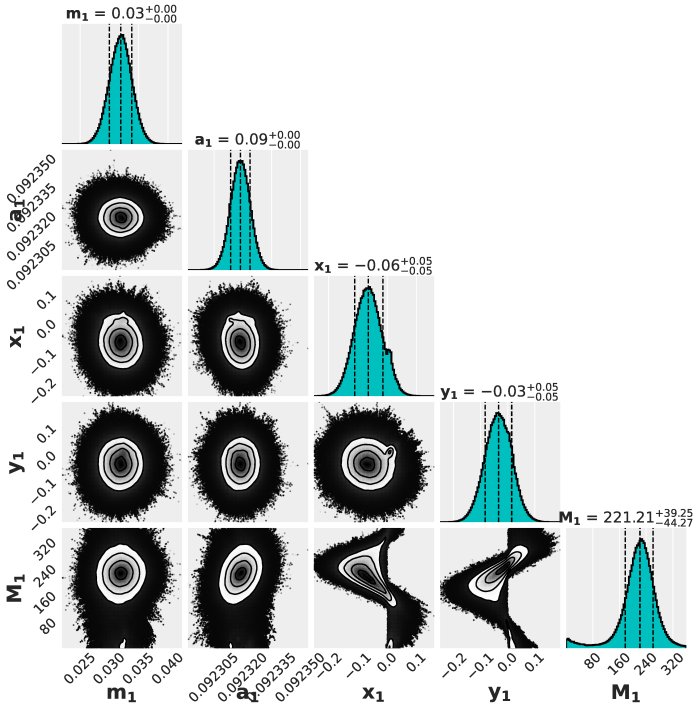}%{../run777777/muAraP1.pdf}
\includegraphics[width=0.5\textwidth]{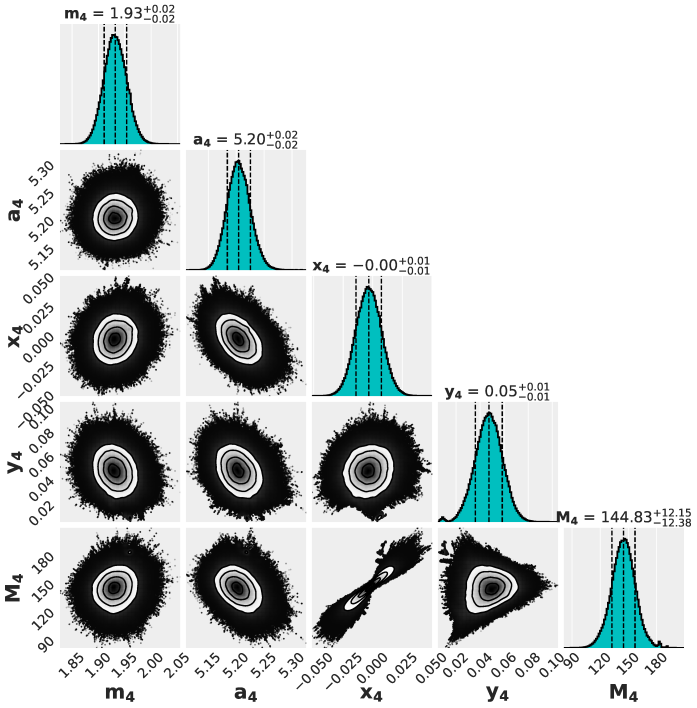}%{../run777777/muAraP4.pdf}
}
}
}
\caption{
One-- and two--dimensional projections of the posterior probability distribution for orbital parameters of the innermost (left column) and the outermost (the right column) planet, respectively. The top row is for the Keplerian model, and the bottom row is for the Newtonian model to data set \Dtwo{}. The parameters are expressed in units consistent with Table~\ref{tab:tab2}. The semi-amplitude $K_i$ is equivalent to the mass $m_i$, and the orbital period $P_i$ is equivalent to the semi-major-axis $a_i$. {The MCMC chain length is 180,000 iterations for each of 384 different instances (walkers) selected in a small ball around a best-fitting solution found with the evolutionary algorithms for the Keplerian model, and 294,000 iterations in each of 176 walkers for the Newtonian model.} Parameter uncertainties are estimated as 16th and 84th percentile samples around the median values at 50th percentile. 
}
\label{fig:fig3}
% figure 3
\end{figure*}

To verify this effect globally in the parameter space, we performed the MCMC sampling with both the Keplerian and Newtonian RV models. The final results for data set \Dtwo{} are illustrated in Fig.~\ref{fig:fig3}. (We skip presentation of the results for \Done{}, since they are very similar). This figure shows one-- and two--dimensional projections of the posterior probability distribution for selected Keplerian (top row) and Newtonian (bottom row) orbital elements obtained for the innermost (left column) and outermost (right column) planet, respectively. The posterior has well defined extrema along all dimensions. We did not notice significant correlations between the displayed parameters, except for ${x,y}$ and ${\cal M}$.

The quality of the best-fit configurations, in terms of RMS $\simeq 3.4$~\ms{}, is also almost the same.  Surprisingly, the posterior distributions are not only very similar to each other, especially if we compare the two-dimensional shape distributions for $x,y$ and ${\cal M}$, but also the eccentricities and orbital angles closely overlap, e.g., the best-fit ${\cal M}_4$ anomaly differs by only $2^{\deg}$ in these models. 

How to interpret this apparent paradox, given the relatively large masses of Jupiter-like companions and their significant, mutual interactions over the observing interval, illustrated in Fig.~\ref{fig:fig2}? A direct comparison of the RV signals may be biased because the accuracy of the formal two-body Keplerian element transformation to Cartesian coordinates is limited to the first order in masses \cite[e.g.][]{Gozdziewski2012}.  However, the representation of the Keplerian initial condition for the $N$-body problem may better fit the data if it is tuned within the parameter uncertainties.  Therefore, given well bounded orbital elements, the MCMC sampling reveals globally similar posteriors for both models. 

We also see the posteriors for the near 2e:1b~MMR pair of a Saturn-Jupiter-like planets exhibiting some significant differences (see on-line Supplementary Material,
Fig.~A1). This can be explained by their relatively shorter periods, covering $\simeq 20$ and $\simeq 10$ times the observational window, respectively, and the 2e:1b~MMR proximity, which strengthens the mutual gravitational interactions.
 
The MCMC experiment implies that, keeping in mind the limitation for representing individual \ics{}, we can still use Keplerian MCMC sampling to efficiently explore the parameter space{, in terms of the posterior distribution}, especially for highly hierarchical configurations with large period ratio. Note that $\Pc{}/\Pd{} \simeq 400$ for \muarae{}. However, parameterization in terms of the $N$-body dynamics is obviously more accurate approach to explain the RV variability when considering individual (local) best-fit models. 

To justify the above explanation, we compared the outcomes of the Keplerian and Newtonian fits for data set \Dtwo{} in Table~\ref{tab:tab2}, and the results are illustrated in the O-C diagram in the right panel in Fig.~\ref{fig:fig2}. This time, the difference between the signals plotted as a red curve in the residuals diagram has much less variability, with the largest differences $\simeq 5$\,\ms{} appearing for epochs without data. 

As noted above, an important feature of the posterior distributions is well bounded parameters for all planets. In particular, the semi-major axes of the middle pair, near 2e:1b-MMR (\araeE{}--\araeB{}) are constrained to $\simeq 0.0015$--$0.002$~au, and for the outermost \araeC{} planet to just $\simeq 0.02$~au , i.e., its orbital period may be determined with the uncertainty of one month {(25--50 times better than with the data in 2006)}. That seems to be quite surprising, since the observational window covers only about 1.5~times the period of this companion. Similarly, the Poincar\'e elements $(x_i = e_i \cos\varpi_i,y_i = e_i \sin\varpi_i)$ of the Saturn- and Jovian planets may be determined to $\pm 0.01$, with uncertainties of {the arguments of pericenter and} the mean anomalies at the osculating epoch $t_0$ on the level of $\pm 15^{\circ}$. {This translates to the mean longitude at the epoch $\lambda_i$ that may be determined to $\simeq 4^{\circ}$.} The eccentricities  in the Keplerian and Newtonian parameterizations (Tables~\ref{tab:tab1}--\ref{tab:tab2}) are  at the $0.05$ level with small uncertainties, as we will show below, may be crucial for maintaining the long-term stability of the system. 

We should also comment on similarities and difference between solutions derived for data sets \Done{} and \Dtwo{} in this work, and with the Keplerian model in \citep{Benedict2022}. 

We obtained very similar eccentricities of the planets, particularly the innermost eccentricity constrained to $e_1 \simeq 0.1$. Given the old age of the star $\simeq 6.7$~Gyr and short orbital period $\simeq 9.64$~days of the {warm} Neptune, its eccentricity might be tidally circularized. We conducted direct numerical integrations of the system with all planets for a few Myr using the SABA$_4$ integrator \citep{Laskar2001} with the step size of $0.5$~days, and we did not detect such a large eccentricity which could be forced by interactions with the outer planets. {Actually, \muarae{}d seems to be a~common example in the known sample of warm Neptunes that exhibit nonzero eccentricity, typically around 0.15 \citep{Correia2020}. They found mechanisms opposing gravitational tides, such as thermal atmospheric tides, evaporation of the atmosphere, and the eccentricity excitation from a distant companion. The later seems to be not the cause of the moderate eccentricity of \muarae{}d, but the presence of atmospheric tides may be sufficient to explain its moderate value.
}

% Most Neptune-mass planets in close-in orbits (orbital periods less than a
% few days) present nonzero eccentricity, typically around 0.15.  This is
% somehow unexpected, as these planets undergo strong tidal dissipation that
% should circularize their orbits in a timescale shorter than the age of the
% system.  In this paper we discuss some mechanisms that can oppose to
% bodily tides, namely, thermal atmospheric tides, evaporation of the
% atmosphere, and excitation from a distant companion.  In the first two
% cases, the eccentricity can increase consistently, while in the last one,
% the eccentricity can only be excited for a limited amount of time (that
% may nevertheless exceed the age of the system).  We show the limitations
% of these different mechanisms and how some of them could, depending on
% specific properties of the observed planetary systems, account for their
% presently observed eccentricities.

The most significant difference between the solutions in \citep{Benedict2022} and in this work is relatively shorter orbital period of \muarae{}c, by $\simeq 100$~days {(yet only $\simeq 2\%$)} in \citep{Benedict2022}. They report this solution as strongly unstable in 100~Kyr time scale, in contrast to our models, which appear safely stable in extended regions of the parameter space, for at least {6.7~Gyr}, as discussed below.

We attempted to address outlying \ucles{} measurements, visible on the right end of the observation window (Fig.~\ref{fig:fig2}). There are systematic deviations from the synthetic model, reaching $\simeq 10$\,\ms{}, and unlikely they can be eliminated with the standard RV ephemeris. The \harps{} and \ucles{} epochs overlap almost throughout the time window, but the \harps{} measurements do not deviate as systematically as the \ucles{} data from the common model. This can be explained by a long-term instrumental \ucles{} effect. In order to account for it, we added a periodic drift to the RV model for the \ucles{} data
$
  \mbox{RV}_{\rm drift}(t) = A \cos( n t + \phi_0 ),
$
where $A$, $n$ and $\phi_0$ are the semi-amplitude, frequency and relative phase of the signal, respectively. 

\begin{figure*}
\centerline{
\hbox{
\includegraphics[width=0.49\textwidth]{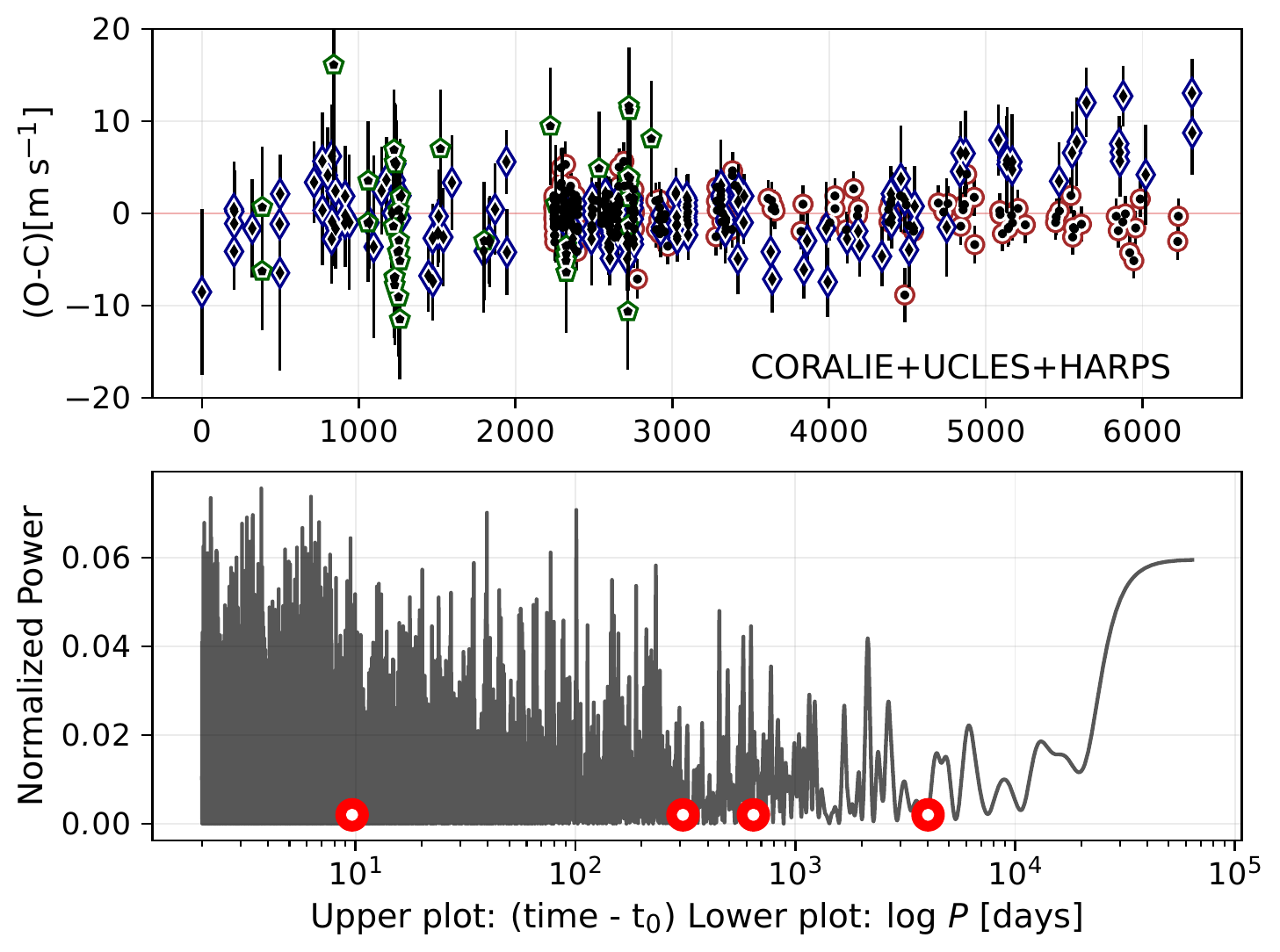}%{../../NewKep/muAraeLomb1Lomb.pdf}
\includegraphics[width=0.49\textwidth]{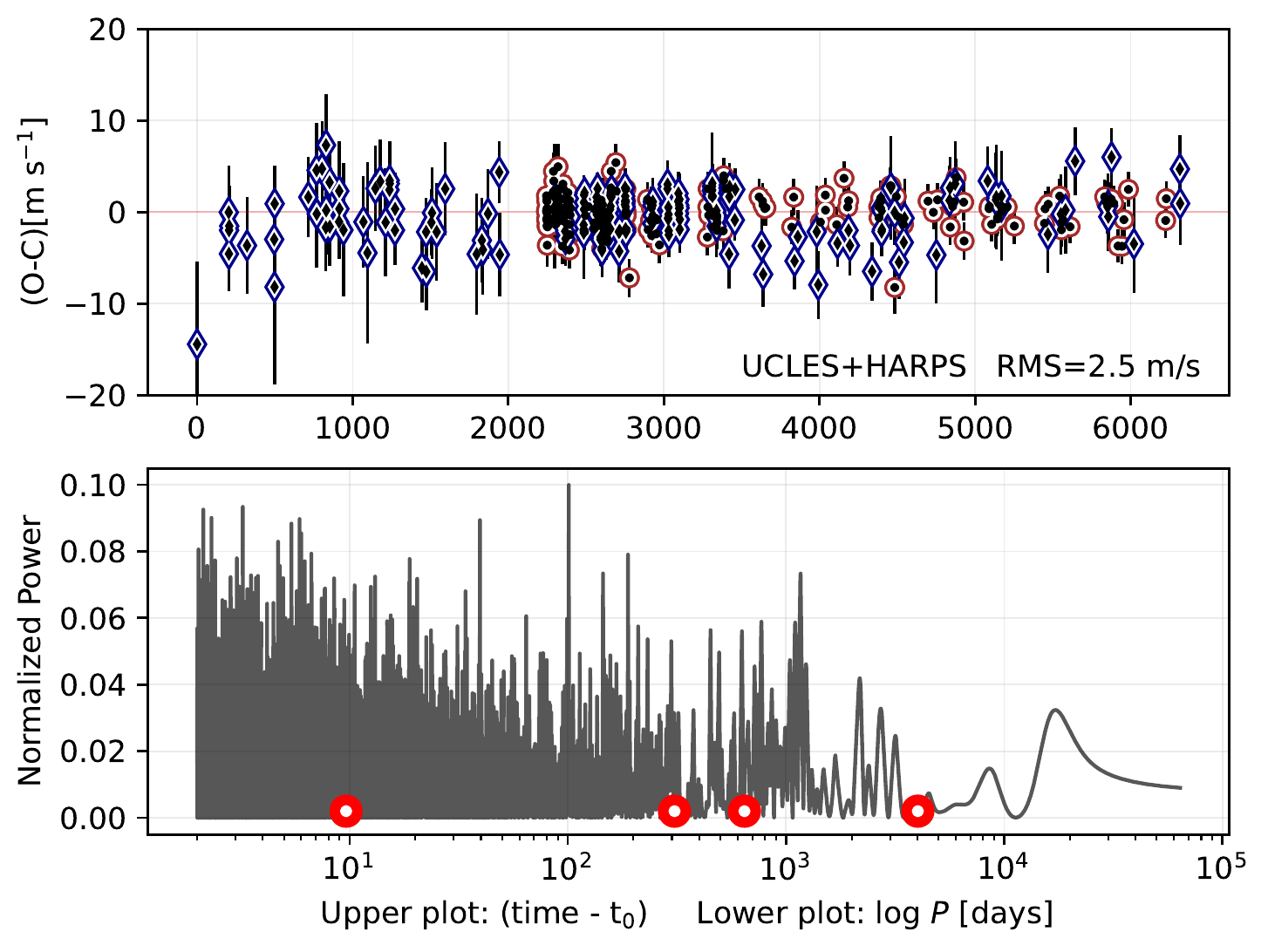}%{../../Doppler5/muArae4XYf.98Lomb.pdf}
}
}
\caption{
{
Lomb-Scargle periodograms for residuals to the Keplerian Fit~IIK in Fig.~\ref{fig:fig2}
(left panel, data symbols are the same as in Fig.~\ref{fig:fig2}) and to the residuals to the best-fitting Keplerian model with a hypothetical, instrumentally induced periodic term $A \cos( n t + \phi_0)$ in the \ucles{} measurements, over-plotted on the RV measurements from \ucles{} and \harps{} spectrometers (right panel).  Brown circles are for \harps{} and blue diamonds are for \ucles{} instrument, respectively. Red filled circles mark the orbital periods of the detected planets.
}
}
\label{fig:fig4}
% figure 4
\end{figure*}

As the result of the MCMC sampling of the Keplerian model with this modification, we show {(O-C) for the best-fit model} in Fig.~\ref{fig:fig4} and a section of the corner plot for the posterior with offsets, error floors, and drift parameters {(on-line Supplementary Material, Fig.~A2)}. Note that in this case we analyzed only the concurrent \harps{} and \ucles{} RV series (data set \Dthree{}). It turns out that the drift component can significantly reduce the \ucles{} outliers. The drift correction reduces the RMS value to $2.5$~\ms{}, which is almost $1$~\ms{} less than the value for the unmodified model. However, the posterior distributions reveal that  the drift's long period $P=2\pi/n \simeq 36$~yrs cannot be meaningfully constrained. Moreover, its half-amplitude $A \simeq 12$--$15$~\ms{} is weakly limited on the right end, and strongly correlated with the RV offset $V_{0,2}\equiv V_{0,{\rm \ucles{}}}$, as it is labeled in the corner plot for the \ucles{} data. At the same time, the orbital parameters have not changed except for the period of $P_4 \simeq (3944 \pm 27)$~days, significantly shorter than $P_4 \simeq 4020$--$4060$~days in our models without drift, but similar to $P_4 \simeq 3947$~days in the solution of \cite{Benedict2022}. 

Given some variability in the residuals to the Keplerian and Newtonian models in Fig.~\ref{fig:fig2}, we analyzed them with the Lomb-Scargle periodogram, in the period window from 2~days to 64,000~days. The results are shown in Fig.~\ref{fig:fig4}. Indeed, the (O-C) in the left panel for the 4-planet model to the data set \Dtwo{} shows some signature of the long-term drift. However, we did not detect any significant peak at the 1\% false alarm probability estimated by the bootstrap method at a level of $\simeq 0.07$. We performed the same test on the residuals to the 4-planet model with the sinusoidal drift. It is clear that the long-term drift period has disappeared, and there are still no significant peaks in the high frequency range.  The (O-C) analysis suggests that we could not detect any significant RV signal that can be attributed to a new planet in the system.

These results are consistent with the conclusions in the work of \cite{Benedict2022}. They did not detect any correlation of the RV variability attributed to the planets with the periodicity of the spectral line profile distortion indicators. They found peaks of the bisector with low significance, around 357--368~days and 497~days, which can be explained by stellar activity.

Since the inclusion of RV drift appears problematic due to the strong $V_{0,2}$--$A$ correlation, and the drift-modified model does not actually qualitatively change the orbital architecture and stability of the system (as justified below), other than shortening the outermost orbital period by $\simeq 2\%$, we have abandoned this model. However, {the likely instrumental} nature and origin of the \ucles{} RV-outliers remains unexplained.

\begin{figure}
\centerline{
%\hbox{
\vbox{
\hbox{\includegraphics[width=0.48\textwidth]{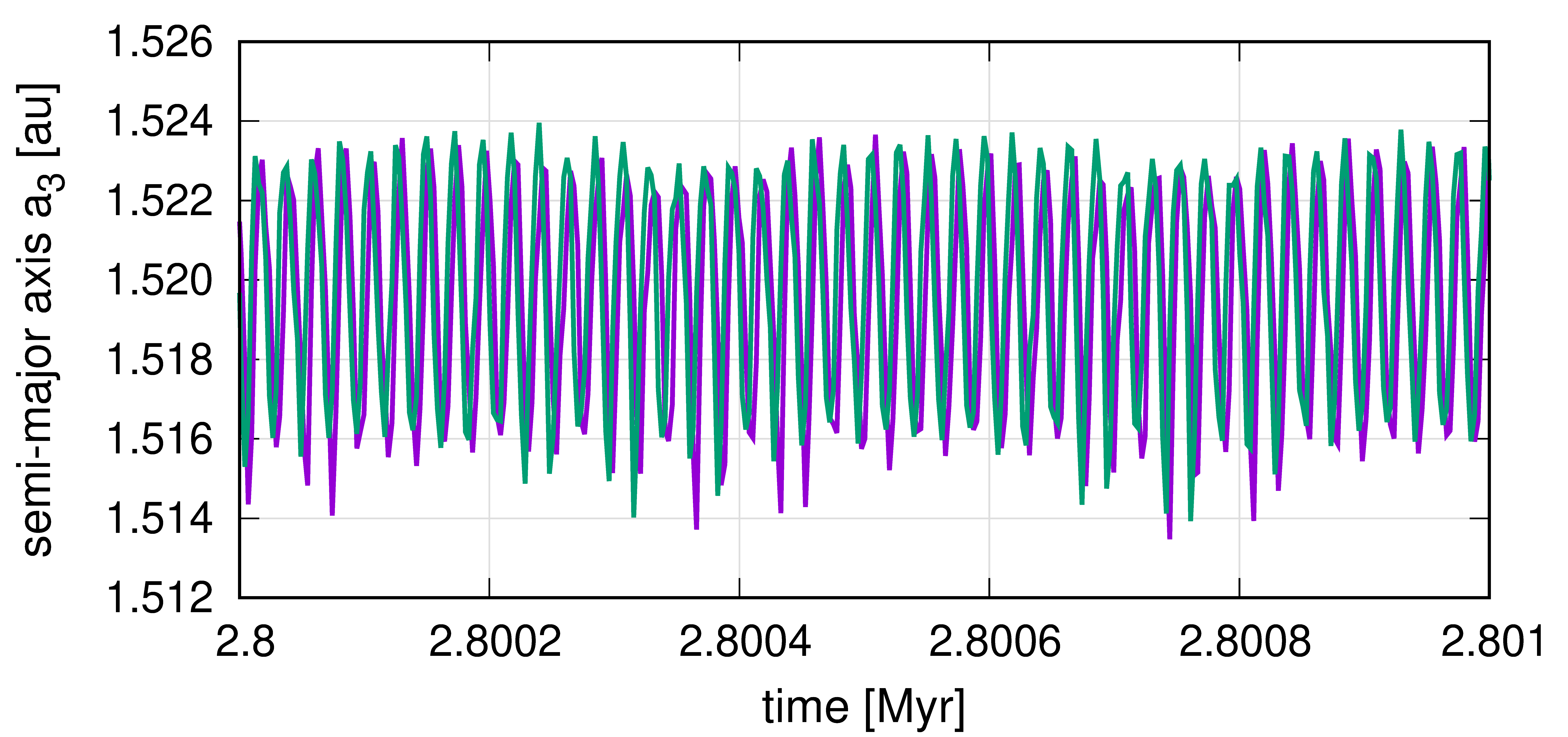}}%{../int.inner/fig3c.png}
\hbox{\includegraphics[width=0.48\textwidth]{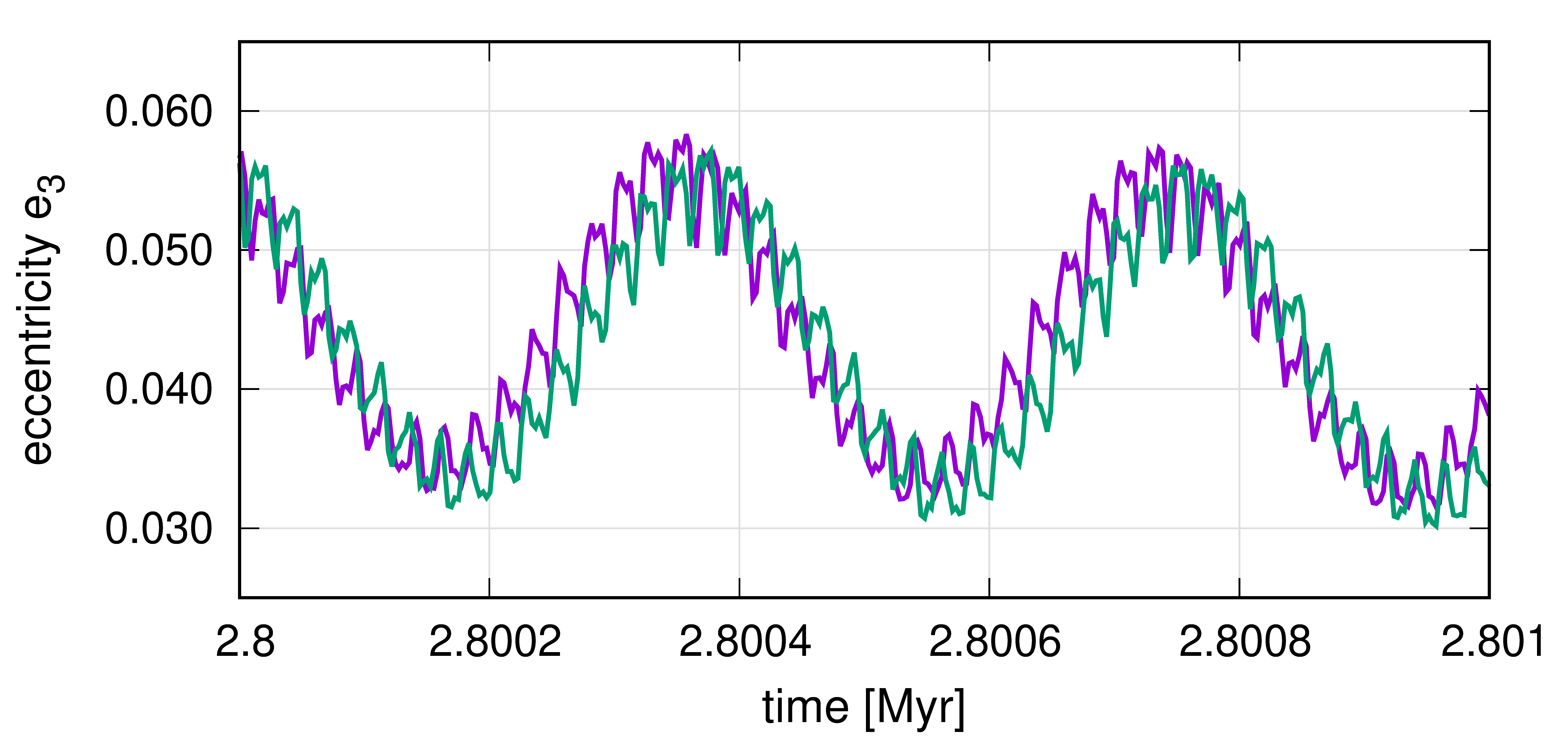}}%{../int.inner/fig3d.png}
}
}
\caption{
Temporal evolution of the osculating semi-major axis ({\em top panel}) and eccentricity ({\em bottom panel}) for planet \muarae{}b in a narrow time window around 2.8~Myr. In each panel, curves with different colour illustrate solutions for two \ics{}, with and without the innermost planet. In the later case, we added its mass to the mass of the star. Elements of the planets included in the integrated system in both experiments are the same (Fit~IN, Table~\ref{tab:tab1}). The systems were integrated with the SABA$_4$ symplectic scheme with the step size of $0.5$~days.
}
\label{fig:fig5}
\end{figure}

\begin{figure*}
\centerline{
\vbox{
\hbox{
\includegraphics[width=0.49\textwidth]{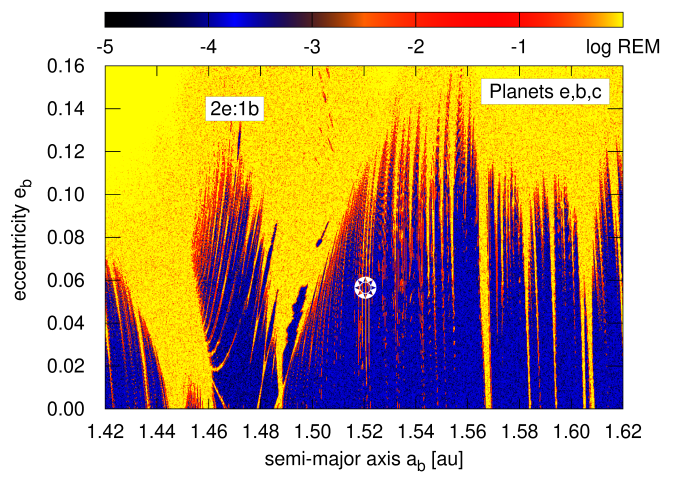}%{../Maps/remcd.png}
\includegraphics[width=0.49\textwidth]{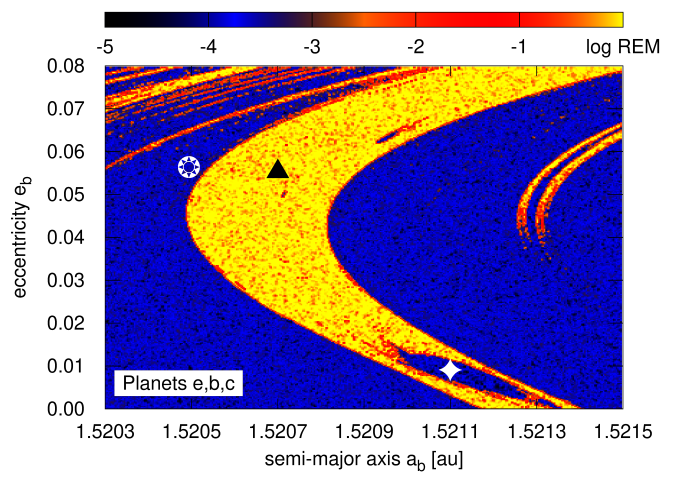}%{../Maps/remcdZoom.png}
}
\hbox{
\includegraphics[width=0.49\textwidth]{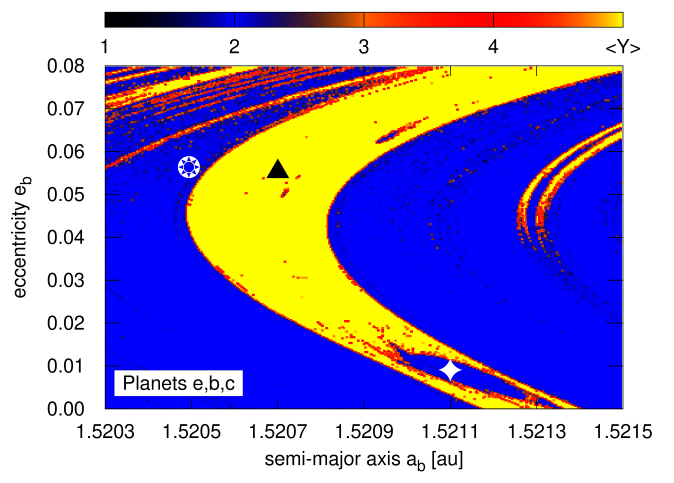}%{../Maps/remcdZoom13.png}
\includegraphics[width=0.49\textwidth]{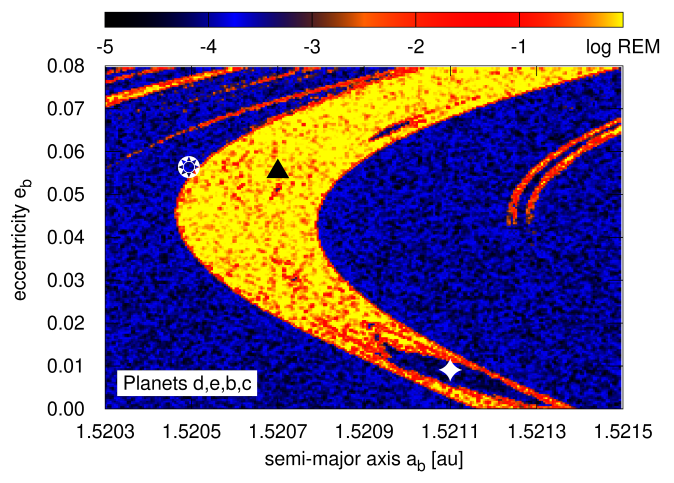}%{../Maps/YcdZoom13.png}
}
}
}
\caption{
Dynamical maps for the best-fitting $N$-body Fit~IN (Tab.~\ref{tab:tab1}) to data set \Done{}. Top-right and bottom panels are for a close-up of the scan shown in top-left panel. The fast indicators $\log |\mbox{\rem{}}| \lesssim -4$ and $\Ym \simeq 2$ characterise regular (long-term stable) solutions, which are marked with black/dark blue colour; chaotic solutions are marked with brighter colors, up to yellow. The integration time of each initial condition is 200~Kyr ($\sim 1.8 \times 10^4 \times \Pc$). Panels in the right column are for the 3-planet model omitting the warn Neptune, {and for the full 4-planet configuration, respectively. The \rem{} indicator was computed with the leap-frog with the step size of 8~days (3-planet map) and 0.33~days, respectively (bottom scan)}. The \megno{} scan (bottom-left panel) was computed for 3-planet model with the Gragg-Bulirsch-Stoer (\gbs{}) algorithm \citep{Hairer1995,Hairer1995b}. The asterisk symbol means the position of the nominal model. Diamond and triangle symbols are mark the \ics{} tested with the direct numerical integrations for {6.7~Gyr, see the text}.  Resolution for the top and bottom-left plots is $640\times 360$ points, {and $360\times 200$ points for the bottom-right scan.}
}
\label{fig:fig6}
\end{figure*}

\begin{figure}
\centerline{
\vbox{
\hbox{\includegraphics[width=0.48\textwidth]{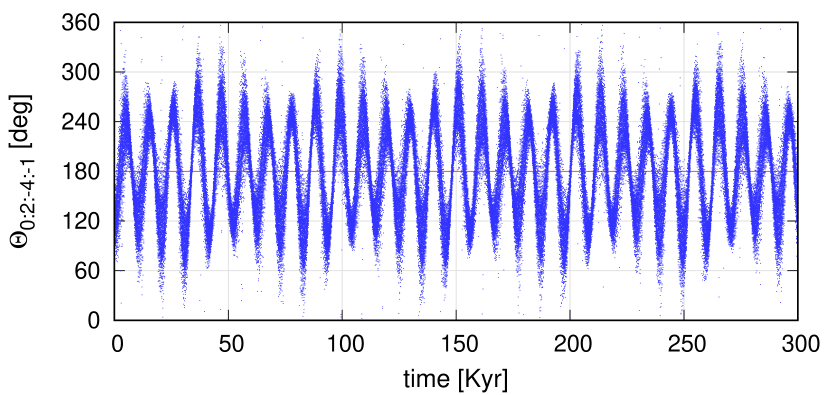}}%{../naff/fig3a.png}
}
}
\caption{
Evolution of a selected critical angle $\theta_{2e:-4b:-1c}$ of the three-body MMR of the outer planets for the initial condition marked in  dynamical maps in Fig.~\ref{fig:fig6} with a white diamond. 
%
% {\em Top panel}: evolution of a selected critical angle $\theta_{2e:-4b:-1c}$ of the three-body MMR of the outer planets for the initial condition marked in a dynamical map in the middle-row panels in Fig.~\ref{fig:fig6} with a white diamond.  {\em Bottom panel}:  temporal evolution of the eccentricities for formally unstable initial condition evaluated for 1~Gyr, marked in the dynamical map with a black triangle.
}
\label{fig:fig7}
\end{figure}

\begin{figure}
\centerline{
\vbox{
\hbox{\includegraphics[width=0.48\textwidth]{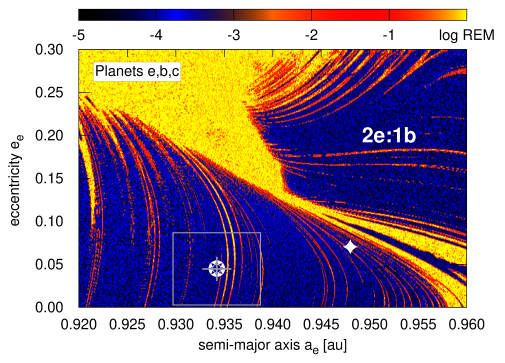}}%{../naff/fig3a.png}
\hbox{\includegraphics[width=0.48\textwidth]{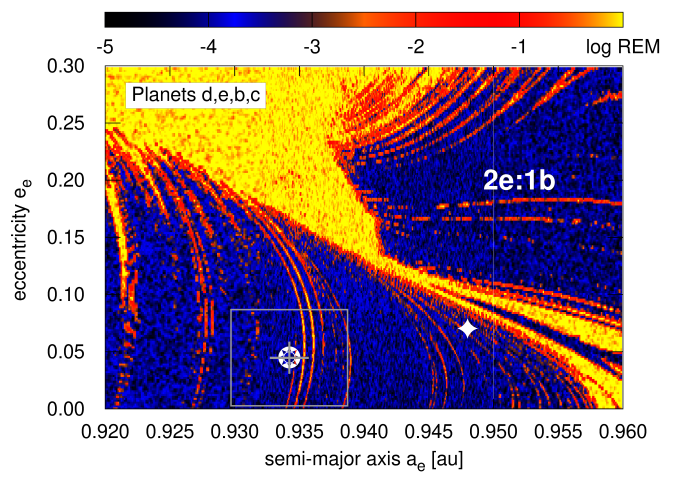}}%{../int.chaotic/fig3b.png}
\hbox{\includegraphics[width=0.48\textwidth]{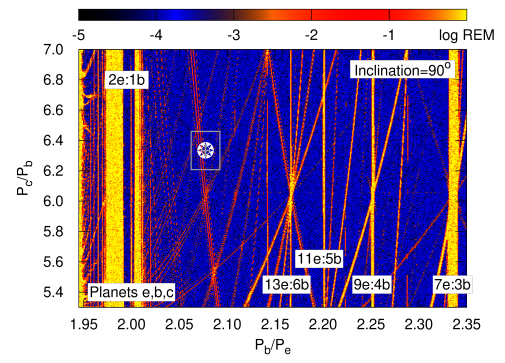}}%{../int.chaotic/fig3b.png}
}
}
\caption{
{
Dynamical maps for Newtonian Fit~IIN (Tab.~\ref{tab:tab2}) to data set \Dtwo{}. The \rem{} with $\log |\mbox{\rem{}}| \lesssim -4$ characterise regular (stable) solutions marked with black/dark blue colour; chaotic solutions are marked with brighter colors. The integration time of each \ics{} is 200~Kyr ($\sim 1.8 \times 10^4 \times \Pc$). The top panel is for 3-planet model (the warm Neptune's mass added to the star mass), and the {middle} panel is for all planets. {The bottom panel is for the \rem{} scan in the plane of Keplerian period ratios, for 3-planet model.} The \rem{} was computed with the leap-frog with the step size of 8~days for 3-planet, and 0.33~days for 4-planet scans, respectively.  The asterisk marks the nominal \ics{} and diamond marks the {\em qualitative} position of the \ics{} in \citep{Pepe2007}, see their Fig.~7. Resolution for the top plot is $1140\times 360$,  $560\times 200$ for the {middle} plot, {and $720\times720$ pixels for the bottom plot}. The cross centered at the \ics{} marks $1\sigma$ error bars ($0.0015$~au, $0.014$) for the $(\ca{},\ee{})$-plane, as in Tab.~\ref{tab:tab2}, and  {for the period ratios $(0.005,0.042)$-plane, respectively.} The gray rectangles are for $3\sigma$ region.
}
}
\label{fig:fig8}
\end{figure}

%_______________________________________________________________________________
%
\section{Long-term stability of the system}
\label{sec:stability}
%_______________________________________________________________________________
%
The well bounded best-fit parameter ranges make is possible to simplify the analysis of the dynamical character of the system. We conducted it with two fast dynamical indicators, the Mean Exponential Growth factor of Nearby Orbits \citep[\megno{}, $\Y$][]{Cincotta2003} and the Reversibility Error Method \citep[\rem{}]{Panichi2017}. These numerical tools are CPU-efficient variants of the Maximal Lyapunov Exponent (\mle{}) that make it possible to detect unstable solutions and visualize the structure of the phase space.

The usefulness of the \megno{} method in analyzing the dynamics of planetary systems with strongly interacting companions has been proven for a long time \citep[e.g.,][and references therein]{Gozdziewski2012}. We have also shown in \citep{Panichi2017} that the \rem{} indicator is not only equivalent to \megno{}, but may be also much more CPU-efficient. Briefly recalling the idea of this algorithm, {computing \rem{} relies in comparing the difference between the Cartesian initial condition $\vec{x}_0$ after integrating it numerically forward and back, for the same number $n$ of time steps $\Delta{}t$, using a time-reversible numerical scheme, to obtain the final state $\vec{x}(\pm n\Delta t)$. Then the \rem{} indicator is
\begin{equation}
\label{eq:rem}
 \mbox{REM} = ||\vec{x}_0-\vec{x}(\pm n\Delta{}t)||.
\end{equation}
}
This difference grows exponentially with integration time for chaotic systems, and at a polynomial rate for regular (stable) configurations. Such a simple algorithm can be implemented with a symplectic discretization scheme. In practice, for systems with small and moderate eccentricities, which \muArae{} systems appear to be, {we use the classic leap-frog algorithm \cite[e.g.][]{Laskar2001} with symplectic correctors of the order~5 \citep{Wisdom2006}, offering numerical accuracy and efficiency comparable to higher order methods \citep{Wisdom2018}, see also \citep{Panichi2017} for details.} As we have shown, in the later paper, this \rem{} algorithm is particularly useful in regions of phase space with predominantly stable solutions and outperforms then any \megno{} variant in terms of CPU-efficiency.  

In this work, to speed up computations, we conducted the numerical simulations using our $\mu${\sc Farm} code parallelized with the Message Passing Interface (MPI). For the numerical integrations of the $N$-body equations of motion {for individual \ics{}}, we used the SABA$_4$ symplectic scheme \citep{Laskar2001} as well as Everhardt's algorithm implemented in the \rebound{} package \citep{Rein2015}.

\subsection{Stability of the model based on data set \Done{}}
We first computed the two-dimensional dynamical maps in the neighborhood of {the Newtonian Fit~IN} in Table~\ref{tab:tab1}, based on the original data set \Done{} from \cite{Benedict2022}. Figure~\ref{fig:fig6} illustrates the $(\ab, \eb)$--plane. In these scans, all other orbital elements are kept at their best-fit values listed in Table~\ref{tab:tab1}. To make possible reproduce the results, we quote exact numerical values of the elements and masses. For each initial condition in the grid, the equations of motion were integrated up to 200~Kyr, corresponding to $\simeq 1.8 \times 10^4 \Pc$. This time interval allows for the detection of short-term chaotic motions for the time scale of the MMRs instability \citep[e.g.][]{GM2018}.

Some of the dynamical maps were computed for 3-planet systems with the most massive planets, omitting the innermost {warm} Neptune. Its very short orbital period of $9.64$~days compared to that one of the outermost planet ($\simeq 4000$~days) causes a huge CPU overhead.  
Before that, we investigated whether the presence of \araeD{} could affect the orbital evolution of the other massive companions and such 3-planet maps. To this end, we numerically integrated the systems described by Fit~IN, with and without the {warm} Neptune, for several Myr, when secular effects may already play a role. Fig.~\ref{fig:fig5} illustrates the resulting osculating semi-major and eccentricity over a narrow time interval around 2.8~Myr for \araeB{} (\muarae{}b). Clearly, the elements span the same ranges and evolve along curves with very similar shapes. Their de-phasing is due to a small change of the mean motion and other elements. The most significant shift can be seen for \araeC{} (\muarae{}c, not shown here), yet its semi-major axes is shifted by $\simeq 0.002$~au, roughly 10~times less than $1\sigma$ uncertainty for this orbital element.

To study whether the innermost planet can be omitted from the system for long-term integrations, \cite{Farago2009} averaged the model for the fast orbiting innermost planet. Obviously, such an analytical model is numerically as CPU efficient, as the 3-planet model. Moreover, they found for the particular \muArae{} case the results from three formulations of the orbital evolution: the exact one, the 3-planet model with omitted warm Neptune, and the 3-planet model with its mass added to the mass of the star lead to barely distinct results.

To test this independently, and without any simplifications of the equations of motion, we used the \rem{} indicator directly and compared dynamical maps for the 3- and 4-planet configurations, respectively, for the same ranges of orbital parameters.

We start with the upper-left panel in Fig.~\ref{fig:fig6} for a relatively broad region of the \ics{} marked with a star symbol. That map was computed without the innermost Neptune, using the leap-frog scheme and a time step of 8~days. A wide structure around $\ab{} \simeq 1.47$\,au on the left of this IC corresponds to the 2b:1c~MMR of the inner pair of Saturn-Jupiter--mass planets. Given the small $1\sigma$ uncertainty $0.001$~au of the nominal semi-major axis, {the} separation of the best-fitting configuration from this MMR is meaningful ({the error bars are smaller than the symbol radius}). Simultaneously, the \ics{} is located between three narrow strips of unstable solutions that may be identified with higher-order resonances. Close-up maps in the remaining panels of Fig.~\ref{fig:fig6} reveal a very close proximity of the \ics{} to one of these strips.

Panels in the bottom row are for the same $(\ab{},\eb{})$-plane, but scanned with $\Y{}$ for the 3-planet model (bottom-left panel) and with \rem{} calculated for the full 4-planet configuration (bottom-right panel), but with a much smaller step size of 0.33~days and lower resolution compared to the 3-planet \rem{}-map computed with the leap-frog step-size 8~days (upper-right panel). Of course, this is forced by the short orbital period of \muarae{}d.  The maps clearly illustrate the one to one results, in a region with weakly unstable configurations and different, very fine dynamical structures. {We may note that the \ics{} is negligibly shifted by $\simeq 10^{-5}$~au with respect to the unstable structure, between the 3-planet and 4-planet scans.} 

While the \rem{} map for three planets was calculated several times faster than the $\Y{}$ map, the full \rem{} calculation for four planets was more than 15 times slower per pixel. Such overhead is acceptable, however, given that the calculations were performed without any simplification of the Newtonian equations of motion.

The detection of fine unstable structures and tiny islands of stable resonances confirms once again a good sensitivity of the \rem{} algorithm for stable and unstable solutions. To show this better, we interpreted the unstable strip structure through the numerical analysis of the fundamental frequencies \citep[NAFF,][]{Laskar2001} of a particular system marked with a white diamond symbol in a small stable island around $(\ab,\eb) \simeq (1.5211\au{},0.01$).  This island is a part of the three-body MMR 2e:-4b:1c structure (one of the strips spanning $\eb \in [0,0.1]$). We plotted evolution of a selected critical angle of this resonance $\theta_{2:-4:-1} = 2\lambda_2 -4 \lambda_3 -\lambda_4 + \varpi_2 +\varpi_3 + \varpi_4$ in Fig.~\ref{fig:fig7}. This critical angle librates with large amplitude around $180^{\circ}$, and the orbital configuration is perfectly stable for at least 1~Gyr, consistently with its location in the stable island. 
 
In contrast, we selected formally unstable \ics{} by shifting the nominal semi-major axis to the right (to the unstable strip) and marked with a black triangle symbol in Fig.~\ref{fig:fig6}. We integrated this \ics{} for 6.7~Gyr with the SABA$_4$ scheme and {for 1~Gyr} with the variable step-size IAS15 integrator.  Also in this case the system does not reveal any signature of geometric instability, in spite of its formally chaotic character in the sense of \mle{} (it is not illustrated here, but we invoke a similar example in Sect.~\ref{sec:inclin}). The width of this third-order MMR is very small, $\Delta{}a_3 \simeq 0.003$~au, and the diffusion is likely so slow that it does not lead to a change or disruption of the system.

We remark here that \cite{Benedict2022} found quite an opposite, catastrophic instability of the system. In their Keplerian solution, $\Pc \simeq (3947 \pm 23)$~days is {apparently} the only significant difference with our fits (Table~\ref{tab:tab1}). The origin of this discrepancy may be a subtly different parameterization of the RV signal. For instance, \cite{Benedict2022} did not fit the jitter uncertainties as free parameters, but tuned it posteriori for each data set to obtain $\Chi \simeq 1$. Moreover, our models yield smaller RMS $\simeq 3.4$\ms{} rather than $\simeq 3.8$\ms{} in the prior work. A shorter period of $\Pc{} \simeq 3947$~days may be pointing to an unstable structure close to $\ac \simeq 5.12$~au (similar to that one visible in the top-left panel in Fig.~\ref{fig:fig10}). We integrated the system with the outermost planet \araeC{} placed in this unstable zone, but the system survived for at least {1~Gyr}. We could not reproduce the strong instability reported in \citep{Benedict2022}, {and we cannot find any convincing explanation of this discrepancy.
}

\subsection{Stability of the Newtonian model based on data set \Dtwo{}}

\begin{figure}
\centerline{
\vbox{
\hbox{\includegraphics[width=0.49\textwidth]{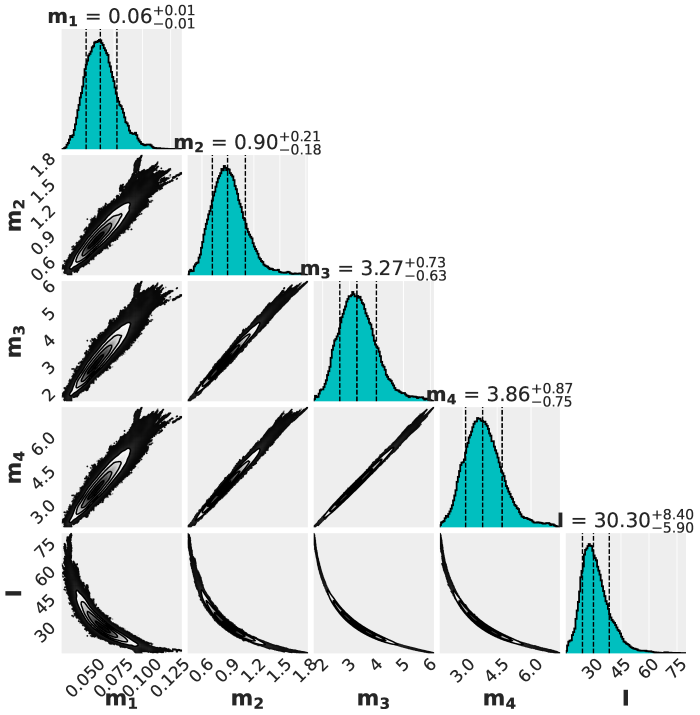}}%{../naff/fig3a.png}
\hbox{\includegraphics[width=0.49\textwidth]{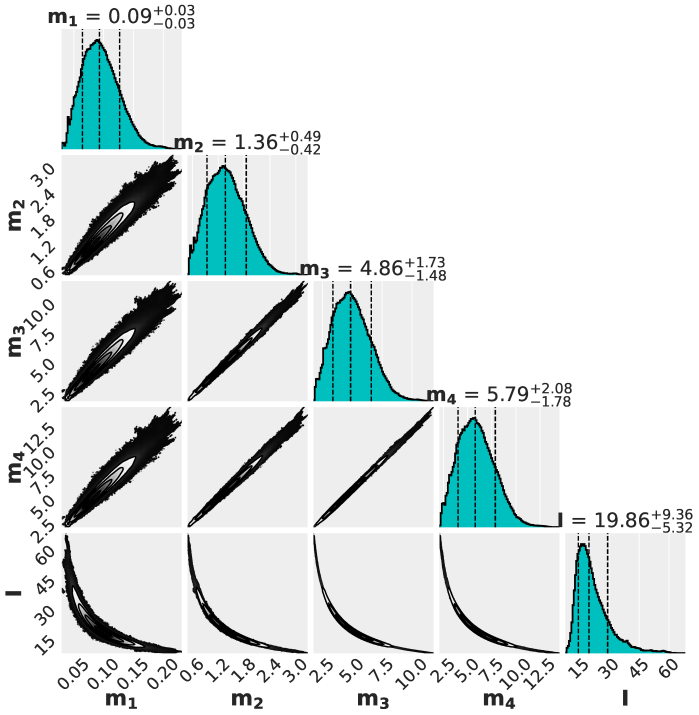}}%{../naff/fig3a.png}
}
}
\caption{
One-- and two--dimensional projections of the posterior probability distribution for the planet masses and the system inclination, illustrating MCMC samples for the Newtonian model fitted with $I$ as a free parameter.
{{\em Upper plot}: The result for data set \Dtwo{}}. The MCMC chain length is 400,000 iterations ($\simeq$ 15 times the greatest auto-correlation time) in each of 144 different instances selected in a small ball around Fit~IIN (Tab.~\ref{tab:tab2}), completed with $I=45^{\circ}$.
{{\em Lower plot}: The result for data set \Dthree{} composed of \harps{} and \ucles{} measurements. The MCMC chain length is 500,000 iterations in each of 144 different instances selected in a small ball encompassing Fit~IIN (Tab.~\ref{tab:tab2}) computed for the osculating epoch  in the middle of the data window and completed with the initial value of $I=45^{\circ}$. }
Parameter uncertainties are estimated as 16th, and 84th percentile samples around the median values (50th percentile) and marked with dashed lines on the 1-dim histograms. Masses $m_{1,2,3,4} \equiv m_{\idm{e,d,b,c}}$ expressed in Jupiter masses, and the inclination $I$ in degrees.
}
\label{fig:fig9}
\end{figure}

\begin{figure*}
\centerline{
\vbox{
\hbox{
\includegraphics[width=0.49\textwidth]{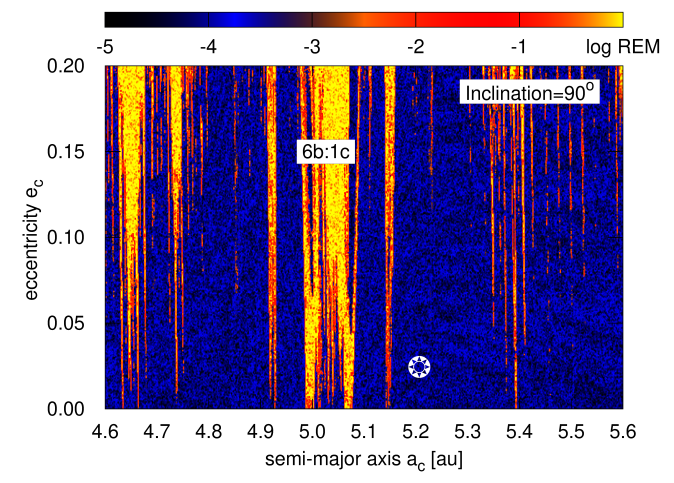}%{../Maps/remEB4NewInc90.png}
\includegraphics[width=0.49\textwidth]{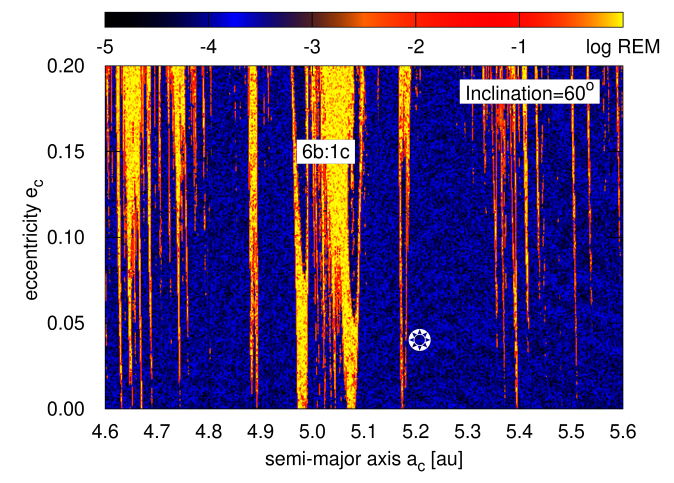}%{../Maps/remInc60.png}
}
\hbox{
\includegraphics[width=0.49\textwidth]{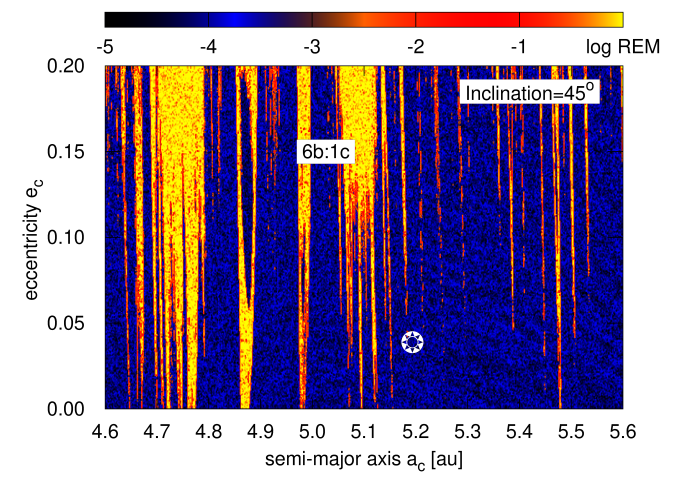}%{../Maps/remInc45.png}
\includegraphics[width=0.49\textwidth]{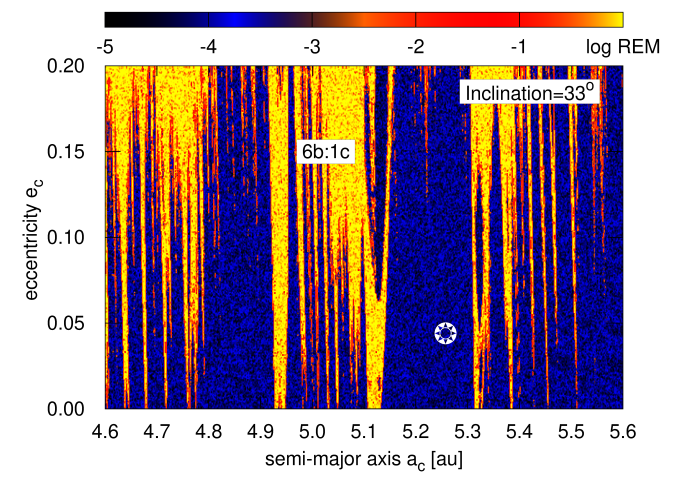}%{../Maps/remInc33.png}
}
}
}
\caption{
Dynamical maps for the best-fitting coplanar $N$-body Fit~IIN (Tab.~\ref{tab:tab2}) to data set \Dtwo{}, extended to the inclination~$I$ space. Subsequent panels are for solutions selected from MCMC samples for the Newtonian model with varied inclination~$I$, illustrated in Fig.~\ref{fig:fig9}, upper plot. The color scale is the same, as in Fig.~\ref{fig:fig6}. The integration time of each initial condition is 300~Kyr ($\sim 2.7 \times 10^4 \times \Pc$); we used the leap-frog scheme with the step size of 8~days. The asterisk symbol means the elements of the nominal fits.  The inclination of the orbital plane is described in the top-right corner of each panel. An approximate position of the 6b:1c~MMR is labeled. Resolution of the plots is $720\times 360$ points.
}
\label{fig:fig10}
\end{figure*}

As mentioned above, we also conducted the GEA and MCMC analysis for data set~\Dtwo{}. The results are very similar to the \Done{} case. However, there are some subtle qualitative changes with respect to the models for \Done{}. The eccentricities of the Jovian planets tend to be systematically even smaller than for the \Done{}--systems. Also the semi-major axes and orbital periods locate the systems in even more ``safe'', stable zone displaced from the 6b:1c MMR by more than 0.1~au, which corresponds to $\simeq 5\sigma$ in terms of the semi-major axis uncertainty.

\subsubsection{The 2e:1b MMR proximity}

\cite{Gozdziewski2007a}, \cite{Pepe2007} and \cite{Farago2009} investigated the proximity of the inner pair \muarae{}e--b to the 2e:1b MMR. In the two later papers, they found the best-fitting model close to the separatrix, unstable zone of this resonance. Contour levels of $\Chi$ encompass both the near-resonance and the resonant configuration \citep[][their Fig.~7]{Pepe2007}. In \citep{Gozdziewski2007a}, we also found that the relative position of the \ics{} and the shape of the 2e:1b resonance in the ($\ab{},\eb{}$)-plane strongly depend on the semi-major axis of \muarae{}c that could be only weakly constrained to $\pm 1300$~days (4~au--7~au) and eccentricity $\ec$ as large as 0.2 at the time.

We can now revisit this issue with a significantly updated Fit~IIN. To do so, we calculated the dynamical maps illustrated in Fig.~\ref{fig:fig8} for the 3-planet (upper panel) and 4-planet (middle panel) configurations, respectively. For the 3-planet model, we added the mass of innermost Neptune to that of the star. It can be clearly seen that the two maps coincide in each detail, and any shift in the position of the \ics{} relative to the fine structures is barely noticeable.

The coordinates of the dynamical maps were chosen to match the NAFF maps in  \cite[][their Fig.~7]{Pepe2007} and in  \citep[][their Fig.~3]{Farago2009}. Since a direct comparison of the maps is not possible, due to changes in elements in the \ics{}, we have marked with a diamond a {qualitative} position of the former initial state relative to the approximate shape of MMR 2e:1b and its separatrix zone.  Clearly, the Fit~IIN is separated from the separatix region by $\simeq 5\sigma$.  This statistically proves that the nominal system is not resonant and is in a safely stable zone. The narrow stripes of unstable motions can be identified with weak, higher-order 3-body MMRs with very long diffusion time scales, similar to the 2e:-4b:-1c MMR analyzed above.

These conclusions can be reinforced with a \rem{} map for the three outer planets in the semi-major axes space, represented in the orbital period ratios $(\Pb{}/\Pe{},\Pc{}/\Pb{})$-plane, as the astrocentric Keplerian representation of the semi-major axes, see the bottom panel of Fig.~\ref{fig:fig8}. Here, we marked $1\sigma$ and $3\sigma$ uncertainties the same as in the previous panels. We computed them based on the MCMC samples. In this map, the 2-body MMRs are marked with vertical (some of them labelled) and horizontal curves. Skewed curves and lines are for 3-body MMRs and could be identified with a method described in \citep{Guzzo2005}. Also this \rem{} map reveals the Fit~IIN safely separated from the 2e:1b MMR by several $\sigma$.

\subsubsection{Stability limits depending on inclination}
\label{sec:inclin}

\begin{figure}
\centerline{
\vbox{
\hbox{\includegraphics[width=0.48\textwidth]{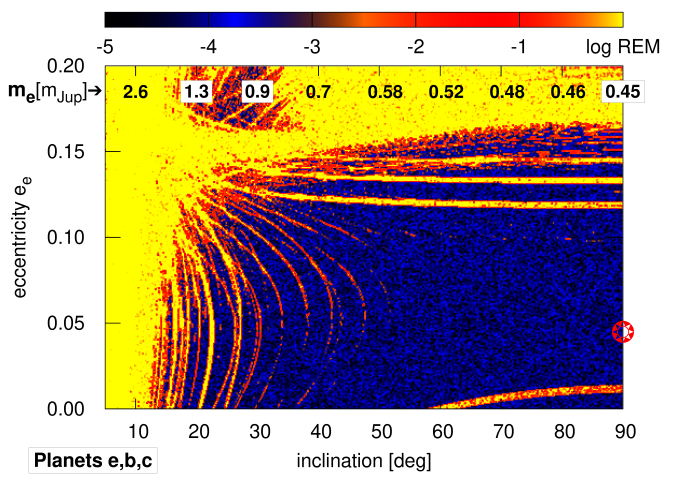}}%{../naff/fig3a.png}
\hbox{\includegraphics[width=0.48\textwidth]{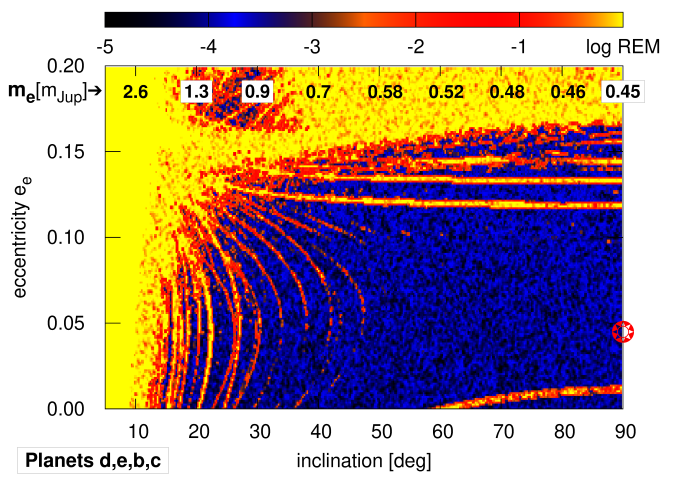}}%{../naff/fig3a.png}
\hbox{\includegraphics[width=0.48\textwidth]{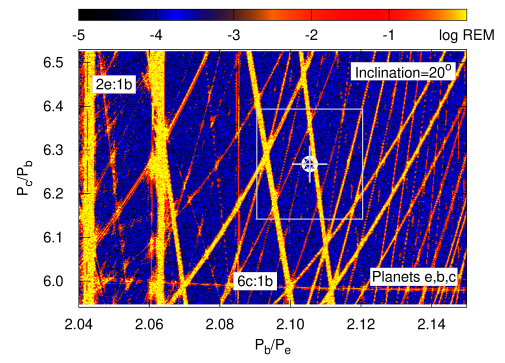}}%{../naff/fig3a.png}
}
}
\caption{
\rem{} dynamical maps for the $N$-body Fit~IIN (Tab.~\ref{tab:tab2}) to data set \Dtwo{}, and planet masses scaled with $m \sin I$ rule. Values $\log |\mbox{\rem{}}| \lesssim -4$  are for long-term stable solutions marked with black/dark blue colour; chaotic solutions are marked with brighter colors, up to yellow. For {the top and middle panels}, the integration time of each \ics{} is 200~Kyr ($\sim 1.8 \times 10^4 \times \Pc$), {and for the bottom panel it is 300~Kyr ($\sim 2.7 \times 10^4 \times \Pc$)}. The top panel is for the 3-planet model with the mass of the warm Neptune added to the mass of the star, and the middle panel is for all 4-planets, respectively. The upper axis marks the mass of $\me{}$ rescaled according to the $m \sin I$ rule. {The bottom panel is for the \rem{} map in the orbital period ratios $(\Pb{}/\Pe{},\Pc{}/\Pb{})$-plane, around the nominal \ics{} found for $I\simeq 20^{\circ}$, close to the posterior maximum in Fig.~\ref{fig:fig9}, bottom plot. Some MMRs are labelled.}  The \rem{} indicator was computed with the leap-frog step size of 8~days for 3-planet and 0.33~days for 4-planet scans, respectively.  Resolution is $640\times 360$, $360\times 200$, and $512 \times 512$ for the subsequent plots, respectively. 
}
\label{fig:fig11}
\end{figure}

Finally, we performed direct MCMC sampling with the inclination added as a free parameter to the Newton co-planar model. As expected, since the RV time series are relatively short covering $\simeq 1.5$ periods of the outermost planet, the inclination may be only weakly constrained in the assumed interval {$[3^{\circ},90^{\circ}]$}. There should be also strong, almost linear correlations between the masses and mass-inclination correlation due to the $m \sin I$ degeneracy. 

However, this intuition seems insufficient in light of the MCMC sampling results for data set \Dtwo{}, illustrated in Fig.~\ref{fig:fig9} (upper plot). It shows posterior histograms for all masses $m_{1,2,3,4}$ and for the inclination~$I$ as a free parameter. In addition to the predicted strong mass-inclination correlation, we found a clear, well-defined posterior maximum for $I\simeq 30^{\circ}$. We tested this effect in multiple MCMC sampling experiments, varying the initial solution and sampling conditions. 

Since, due to parameter correlations, the estimated auto-correlation time is as many as $25,000$ iterations, we sampled up to $400,000$ steps for each of 144~walkers, corresponding to $15-20$ auto-correlation times. As a starting point for the sampling, we took Fit~IIN in Table~\ref{fig:fig2} supplemented with $I=20^{\circ},45^{\circ},60^{\circ}$ and $75^{\circ}$, respectively. Interestingly, in all cases, regardless of the initial~$I$, the extremum is robust and occurs around $I  \simeq (30^{\circ}\pm 10^{\circ})$. At the same time, we monitored the RMS $>3.4$\,\ms{} for best-fitting solutions, which rises significantly to RMS $\simeq 3.6$--$3.8$\,\ms{} below $I>30^{\circ}$. This means that the RV data predicts all planetary masses safely below the brown dwarf limit, i.e., the physical masses can be at most 2--3 times the minimum masses. 

To assess the statistical significance of this result, we computed the Bayesian information criterion (\bic{})  defined as \citep[e.g.][]{Claeskens2008}
\[
\mbox{\bic{}} = p \ln \Nm{} - 2 \lnL_{\idm{max}},
\]
for the Newtonian model, for the edge-on system with $I=90^{\circ}$ and for a model with variable $I$, with  $p=26$ and $p=27$ of free parameters, respectively; $\Nm{}=411$, and $\lnL_{\idm{max}}$ is the value of $\lnL$ evaluated at the posterior extremum. For the two models, we found $\lnL_{\idm{max}}(\vec{\theta},I=90^{\circ})=-987.07$ and $\lnL_{\idm{max}}(\vec{\theta,I})=-987.7$, respectively, hence $\mbox{\bic{}}(\vec{\theta},I=90^{\circ})=2130.62$, and $\mbox{\bic{}}(\vec{\theta},I)=2137.96$, respectively. Therefore 
\[
\Delta\mbox{\bic{}} = 
\mbox{\bic{}}(\vec{\theta},I=90^{\circ})- \mbox{\bic{}}(\vec{\theta},I) \simeq -7 <2,
\]
indicating that there is no evidence of the model with free inclination against the edge-on model with a smaller value of $\mbox{\bic{}}$, see \citep{Claeskens2008}. However, if we apply the second-order Akaike information criterion (\aic{}) for small sample sizes ($\Nm{}/p \simeq 15 <40$),
\[
\mbox{AIC} = 2 p + 2(p+1)/(\Nm{}-p-1) - 2 \lnL_{\idm{max}},
\]
then $\Delta{}\mbox{AIC}<2$ for the two concurrent fit models, and that the candidate model is indicated almost as good as the best edge-on model \citep{Claeskens2008}. We consider this as a marginal indication of the significance of the inclined model, which needs to be addressed with longer RV time series.

Furthermore, we examined this effect for the \Dthree{} data set, consisting of only the most accurate \harps{} and \ucles{} RVs, and also changed the osculating epoch of the Newtonian model to the middle of the RV time series. In this experiment, we also increased the number of iterations to $500,000$ steps for each of the 144 walkers. As a starting \ics{}, we chose Fit IIN from Tab.~\ref{tab:tab2} with an initial value of $I=45^{\circ}$, but without any prior tuning of this solution. The results are shown in Fig.~\ref{fig:fig9}, lower plot. In this case, the posterior distribution is shifted toward $I=20^{\circ}$. This may further indicate a systematic but weak dependence of the Newtonian model on the inclination, which is also sensitive to the RVs changes.

\begin{figure}
\centerline{
\hbox{\includegraphics[width=0.48\textwidth]{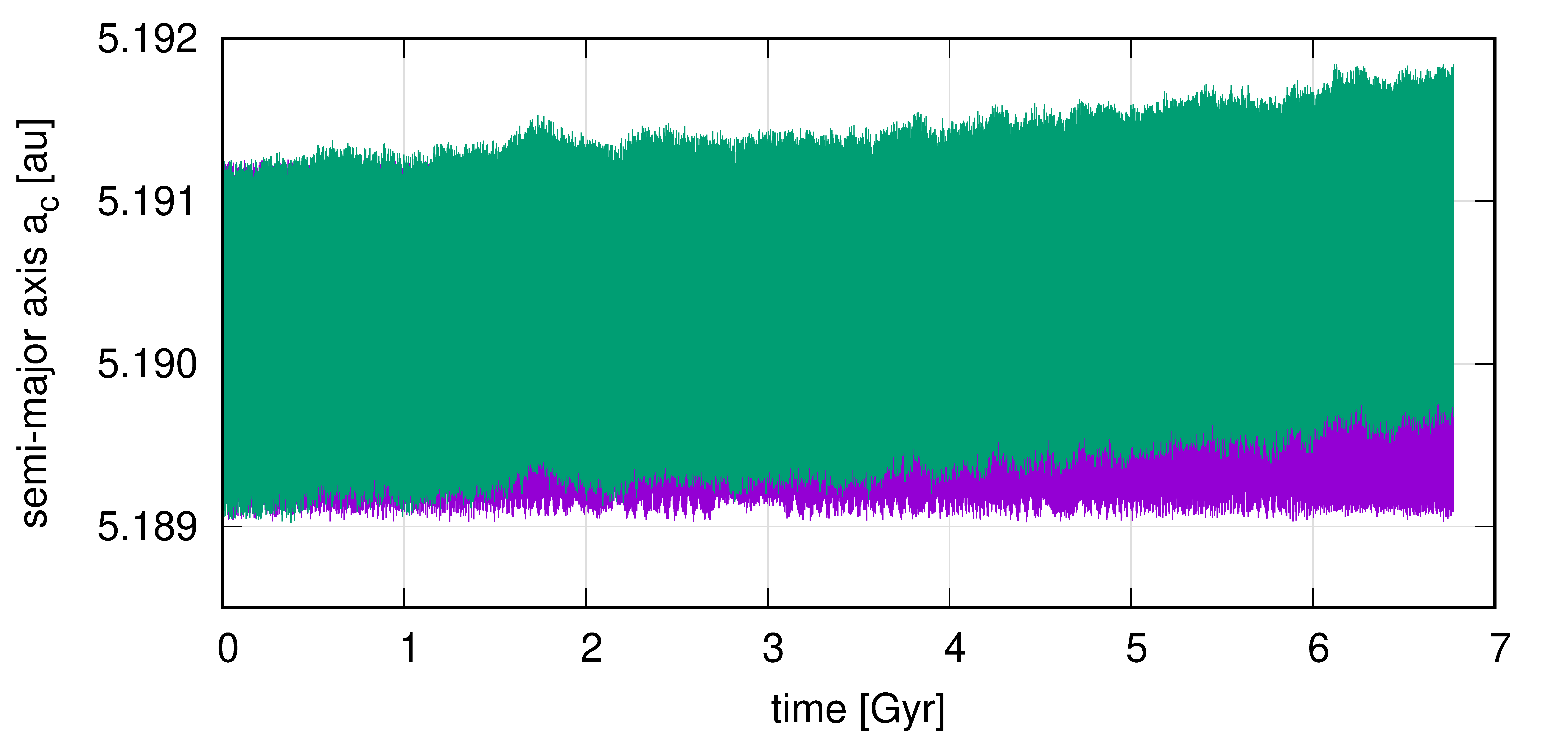}}
}
\caption{
Time-evolution of the semi-major axis of \muarae{}c (drawn in magenta) for the nominal system marked in the lower panel of Fig.~\ref{fig:fig11} with the star symbol, and for a system shifted to the nearby unstable 3-body MMR structure (drawn in green) respectively. The configurations were integrated with SABA$_4$ scheme and the step size of 16~days for 6.7~Gyrs. Chaotic diffusion for the unstable resonant model is apparent. (The \ics{} for this model is given in the Supplementary Material on-line).
}
\label{fig:fig12}
\end{figure}

The stability zone and fine unstable structures {for inclined co-planar systems} are illustrated in dynamical maps in the $(\ac,\ec)$-plane (Fig.~\ref{fig:fig10})
constructed for different inclinations of the co-planar system.  We selected the best-fitting solutions from the MCMC samples with lowest RMS $\simeq 3.35$\,\ms{} detected, and close to particular, a'priori fixed inclinations. Subsequent panels are for such best-fitting models with the inclination equal to $I=90^{\circ}$ (the nominal Fit~IIN in Tab.~\ref{tab:tab2}), $I=60^{\circ}$, $I=45^{\circ}$, and $I=33^{\circ}$, respectively. In the later case, the planet masses are twice as large as in the nominal, edge-on system. Moreover, the orbital elements selected from the MCMC samples are slightly different, thus introducing variability consistent with parameter uncertainties to the elements behind the map coordinates.

To effectively illustrate the region of stability with respect to $I$ in a more global way, we scaled the minimal masses in Fit~IIN according to the minimum mass rule $m_i \sin I=\mbox{const}$, recalling the mass-inclination correlation. We then calculated the dynamical maps in the $(I,\ee{})$ plane (Fig.~\ref{fig:fig11}). For reference, the second upper $x$ axis in these maps is for the mass of \muarae{}e scaled with $I$. 

Although, as we have shown, the influence of the warm Neptune is negligible for the dynamical evolution of the outer planets when their masses are minimal, this may not be the case for small inclinations. We therefore calculated two versions of the \rem{} maps, for three- (top panel) and four-planets (middle panel), respectively (the later with lower resolution to save CPU time). It can be clearly seen that in the range of $I \in [5^{\circ},90^{\circ}]$, which covers the variation of masses spanning one order of magnitude, all, even very fine features of the phase space remain the same.

Finally, we constructed a \rem{} map in the orbital period ratios plane shown in Fig.~\ref{fig:fig11} (bottom panel) around $I=20^{\circ}$, similar to the scan in Fig.~\ref{fig:fig8}. In this case, the masses of the planets are $(1.34,4.91,5.85)\,\mJ{}$, i.e., the minimum masses scaled by a factor~3. We integrated each point for 300~kyr forward and back with the leap-frog scheme and the step size of 8~days. The \ics{} is located in a denser network of 2-body and 3-body MMRs, but still well separated from the 2e:1b MMR. We can also observe the high sensitivity of \rem{} to interacting MMRs, indicated by in their regions of overlap (crossings).

Since the \ics{} is very close to an unstable 3-body MMR, we performed a comparative integration of the nominal system and a configuration slightly shifted so that it is located in this nearby unstable MMR region (yellow strip in the lower panel of Fig.~\ref{fig:fig11}). We used the SABA$_4$ scheme and the step size of 16~days, keeping the energy integral to $10^{-10}$ on the relative scale. In both cases, the system survived integrations for the lifetime of the star (6.7~Gyr). Such narrow chaotic 3-body MMRs, similar to that one analysed in Fig.~\ref{fig:fig4} do not appear ``dangerous'' for the long-term stability. The chaotic configuration reveals only weak diffusion of $\ac{}$ and $\ec{}$. This is illustrated in Fig.~\ref{fig:fig12}.

The general conclusion of this experiment is a relatively wide stable zone preserved despite the enlarged minimal masses of the planets 2-3 times. The limit of stable solutions for $I=15^{\circ}$--$20^{\circ}$ roughly coincides with the shape of statistically detected posterior extremum for $I=30^{\circ}$ (data set \Dtwo{}) and {$I=20^{\circ}$ for data set \Dthree{}}, as we found with the MCMC sampling. Systems with the most probable inclinations $I=60^{\circ}$ in purely random sample would be in the middle of a broad, stable zone. Such the likely inclination increases the planet masses by only $15\%$.

Moreover, the clear posterior maxima for $I\simeq 30^{\circ}$ and $I\simeq 20^{\circ}$ found here (still, in the stable zone) may confirm the marginally detected bias toward small inclinations of multiple systems, investigated with the \hst{} astrometry in \citep{Benedict2022}. We should also note that for \muArae{} very small inclinations $I \lesssim  10^{\circ}$ can apparently be ruled out on both statistical as well as on dynamical grounds.

%  
%_______________________________________________________________________________
%
\section{Possible debris disks and smaller planets}
%_______________________________________________________________________________
%
\label{sec:debris}

Based on the updated, {rigorously stable and well constrained orbital solutions} collected in Table~\ref{tab:tab2}, we simulated the dynamical structure of hypothetical debris disks in the system. In the large ``gap'' between the two outer planets, at $\simeq 1.52$~au and $5.2$~au, respectively, we can predict orbitally stable objects with masses that are below the present detection levels. This region may be an analogue of the Main Belt in the Solar System, given the striking similarity of the orbits of the Saturn- and Jupiter-mass planets to those of Mars and Jupiter. The second debris disk, located beyond the orbit of the outermost Jovian planet (\araeC{}), may be similar to the Kuiper Belt. There is also free space between the two innermost planets that may contain Earth-mass objects, in the wide free space extending for $\simeq 0.9$~au between the orbits.

We could try to recover the structure of the phase space using fast indicators, in the form of the dynamic maps shown earlier in Fig.~\ref{fig:fig6} and \ref{fig:fig10} for the planets. However, such maps constructed for fixed orbital phases of test particles {permitted to vary freely} may reveal an incomplete picture. The stability of a free test body in a system depends not only on its semi-major axis and eccentricity $(a_0,e_0$), but also on its relative orbital phase with respect to massive planetary perturbers.

To circumvent this limitation, we introduced a concept of the so-called $\Ym$-model (or $\Ym$-disk) \citep{GM2018}. We assume that the massive planets form a system of primaries in safely stable orbits robust to small perturbations. Then we inject bodies with masses significantly smaller than masses of the primaries on orbits with different semi-major axes and eccentricities spanning the interesting region, and the orbital phases selected randomly. Next, we integrate numerically the individual synthetic configurations and determine their stability with the \megno{} aka $\Ym$ fast indicator. For this experiment \megno{} is preferable over \rem{} since we may expect that most of the orbits are unstable. As soon as \megno{} reaches a value $\Ym \simeq 5$, sufficiently different for $\Ym \simeq 2$ for stable solutions, we can stop the integration, thus saving CPU-time.  We explain in detail the method and calibration experiments spanning orbital evolution of debris disks in the massive four-planet \hr8799{} system for up to $70$~Myr in \citep{GM2018}. A comparison of the results of direct numerical integrations with the outcomes of the $\Ym$-model confirms that these two approaches are consistent one with the other. Yet the $\Ym$-disk method is CPU-efficient and therefore makes it possible to obtain a clear, quasi-global representation of the structure of stable solutions. This algorithm is especially effective for strongly interacting systems.  

\begin{figure*}
\centerline{
\hbox{
\vbox{ % 0.48
\hbox{\includegraphics[width=0.37\textwidth]{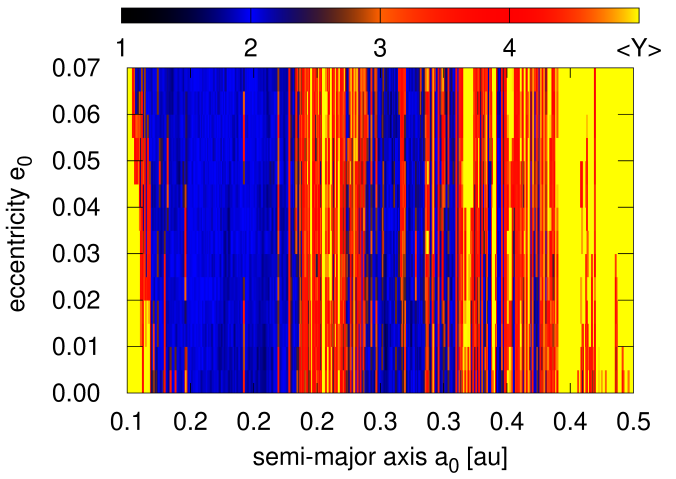}}%{../Maps/YEarth33.png}
\hbox{\includegraphics[width=0.37\textwidth]{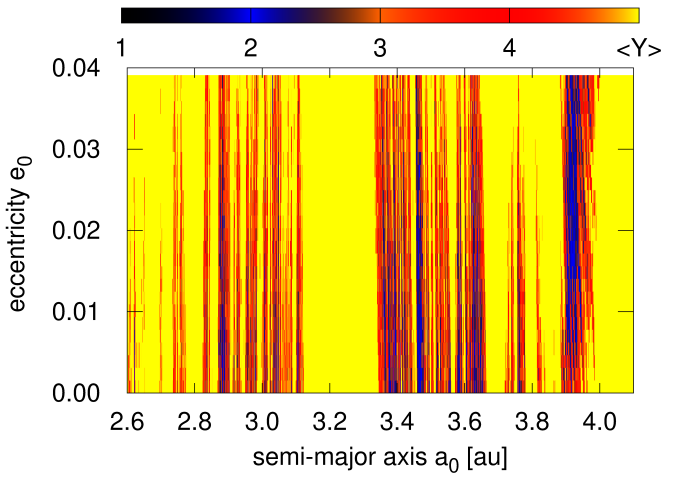}}%{../Maps/YEarth222.png}
\hbox{\includegraphics[width=0.37\textwidth]{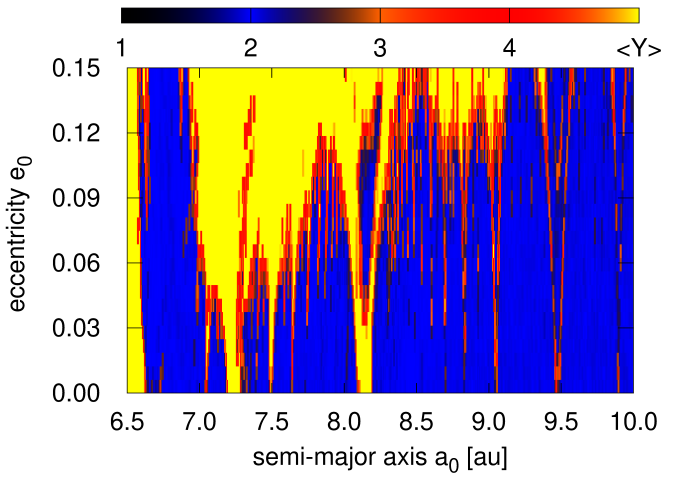}}%{../Maps/YEarth55.png}
}
}
\vbox{
\hbox{\includegraphics[width=0.6\textwidth]{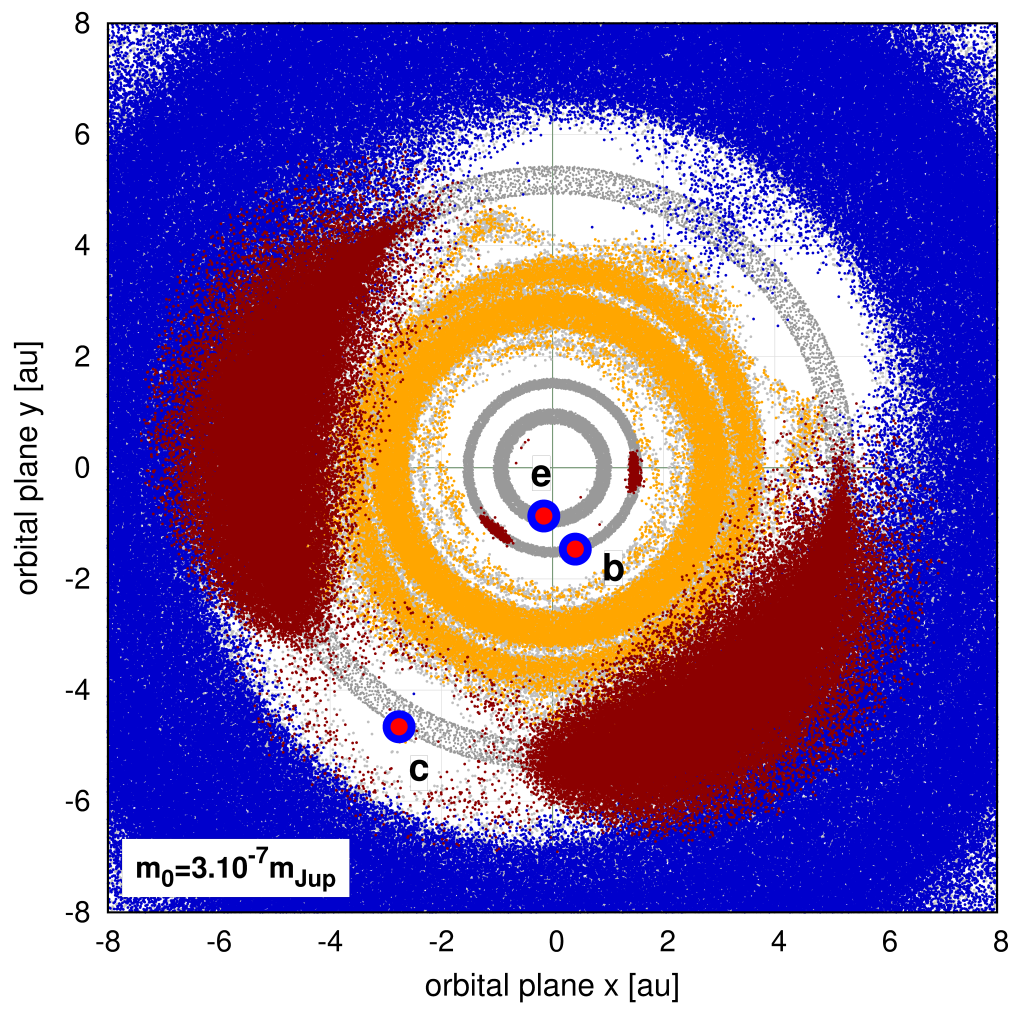}}%{../disks/DiskXY.png}
\hbox{\includegraphics[width=0.6\textwidth]{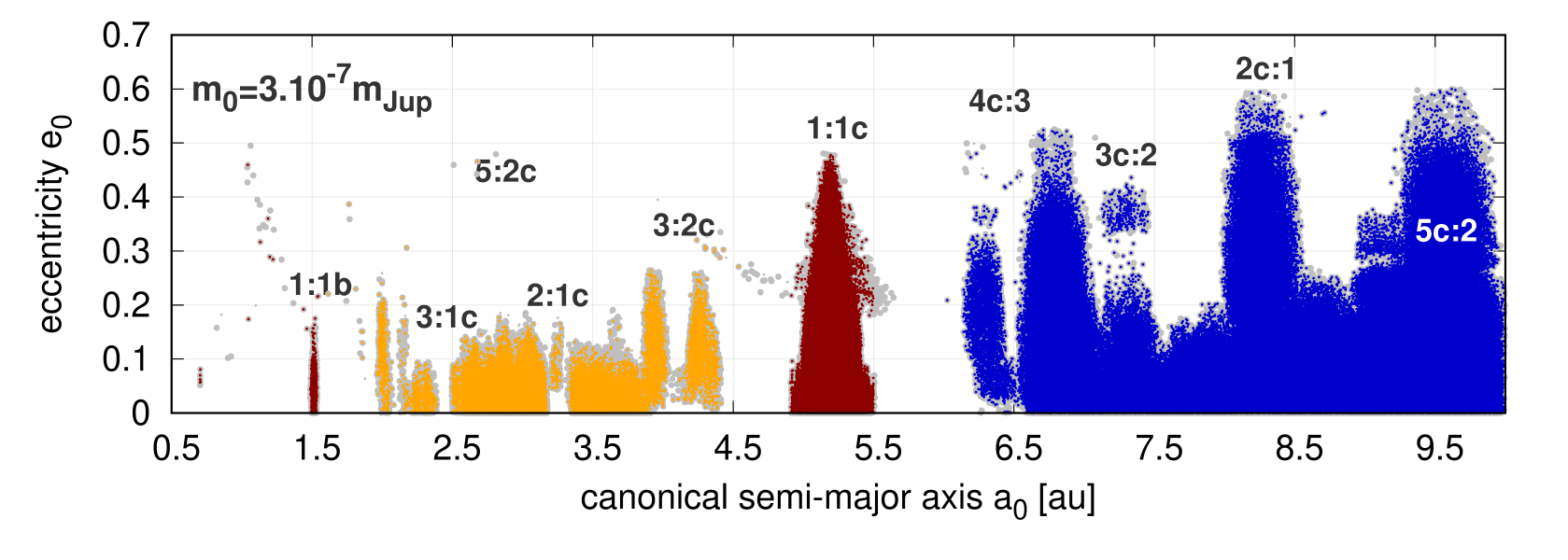}}%{../disks/DiskAE.png}
}
}
\caption{ 
{\em Left column}: dynamical maps for a test particle (Vesta-like asteroid with the mass of $3\times 10^{-7}\mJ$) in three distinct regions between the planets. The \megno{} $\Y \sim 2$ indicates a regular (long-term stable) solution marked with black/dark blue colour, $\Y$ much larger than $2$, up to $\gtrsim 5$ indicates a chaotic solution (yellow). The integration time of each initial condition is $10^5$~yr ($\simeq 10^4 \times \Pc$). 
{\em Top-right panel}:
Debris disks in the \muarae{} system {revealed by $\simeq 10^6$} stable orbits with $|\Ym-2|<0.007$ gathered in the $\Ym$-disk simulation. They are illustrated as a snapshot of astrocentric coordinates $(x, y)$ at the initial epoch $t_0$. Colors of test particles injected with random elements, $a_0 \in [0.9,10]$~au and $e_0 \in [0,0.6])$ into the system of three outer planets, correspond to their dynamical status marked also in the panel with orbital elements, below. The initial positions of the planets are marked with filled circles. Gray rings illustrate their orbits integrated in a separate run for 0.2~Myr.
{\em Bottom-right panel}: the orbital structure of hypothetical debris disks in the system, in terms of canonical Jacobi elements in the $(a_0,e_0)$-plane. Some two-body, lowest order MMRs with the planets are labeled, and stable orbits in their regions are marked with different colors, consistent with a snapshot of these stable solutions in the above panel.
}
\label{fig:fig13}
\end{figure*}

To conduct the $\Ym$ simulations, we chose Fit~IIN {located in a wide zone of stable motions. Such a ``safe'' neighborhood is important for the $\Ym$-model, since the system is not prone to small perturbations exerted by the probe objects -- we integrate numerically the orbits of all bodies.} Again, {since we focus on the space beyond the orbit of \muarae{}e, $a_0\simeq 0.9$~au, we omitted the innermost planet influence for the Main Belt and Kuiper Belt disks, to improve the CPU performance. However, the effect of the innermost planet} was included in the simulation of the dynamical map for the inner zone between the {warm} Neptune and the Saturn-like planet (upper left panel in Fig.~\ref{fig:fig13}).

We considered three types of probe objects in different mass regime: Vesta-like asteroids with a mass of $3\times 10^{-7}\,\mJ$, massive Earth-like planets with a mass of $10^{-2}\,\mJ$, and super-Earths with a mass of $3\times 10^{-2}\,\mJ$ (equivalent to $\simeq 10$ Earth masses, in a sub-Neptune mass range). The RV amplitude of the later objects would be on the level of $2-3$\,\ms{}, relatively easily detectable with the present RV measurements accuracy. Also, in that case we set the system inclination $I=60^{\circ}$ to enhance the mutual gravitational influence between the planets and the test objects. In all experiments, the probe object interacts gravitationally with the three most massive planets.
 
To calculate the $\Ym$ values for the synthetic systems, we integrated the $N$-body equations of motion and their variational equations with the \gbs{} integrator \citep{Hairer1995,Hairer1995b} for $\simeq 10^5$~yrs. Such an interval covers $\simeq 10^4$ orbital periods of outermost planet, which makes it possible to detect unstable motions associated with strongest two-body and three-body MMRs. This integration time is also consistent with the the typical characteristic time-scale required to achieve $\Ym$ convergence for a stable configuration. {The \gbs{} integrator is the best choice in the case of collisional dynamics that is frequently expected in this setup.}

\begin{figure*}
\centerline{
\vbox{
\hbox{ % 0.48
\hbox{\includegraphics[width=0.49\textwidth]{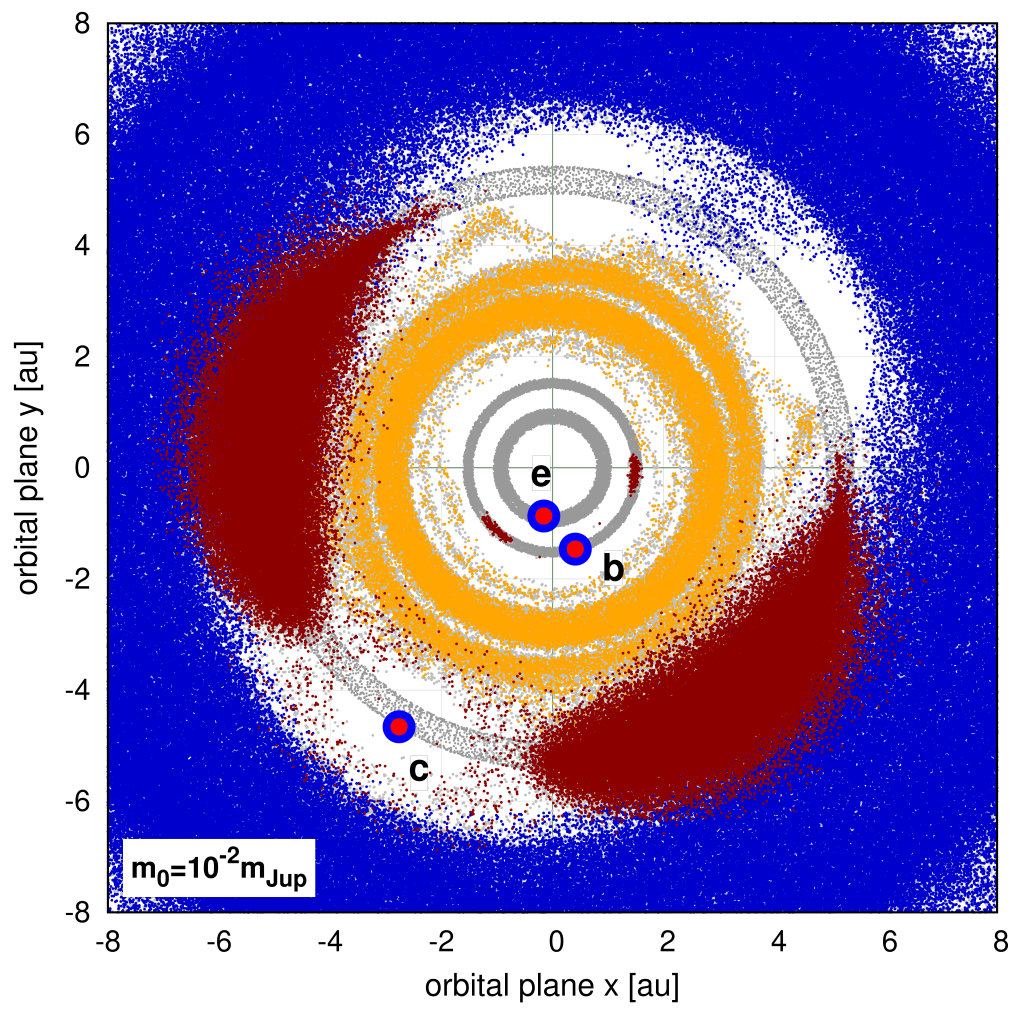}}%{../Maps/YEarth33.png}
\hbox{\includegraphics[width=0.49\textwidth]{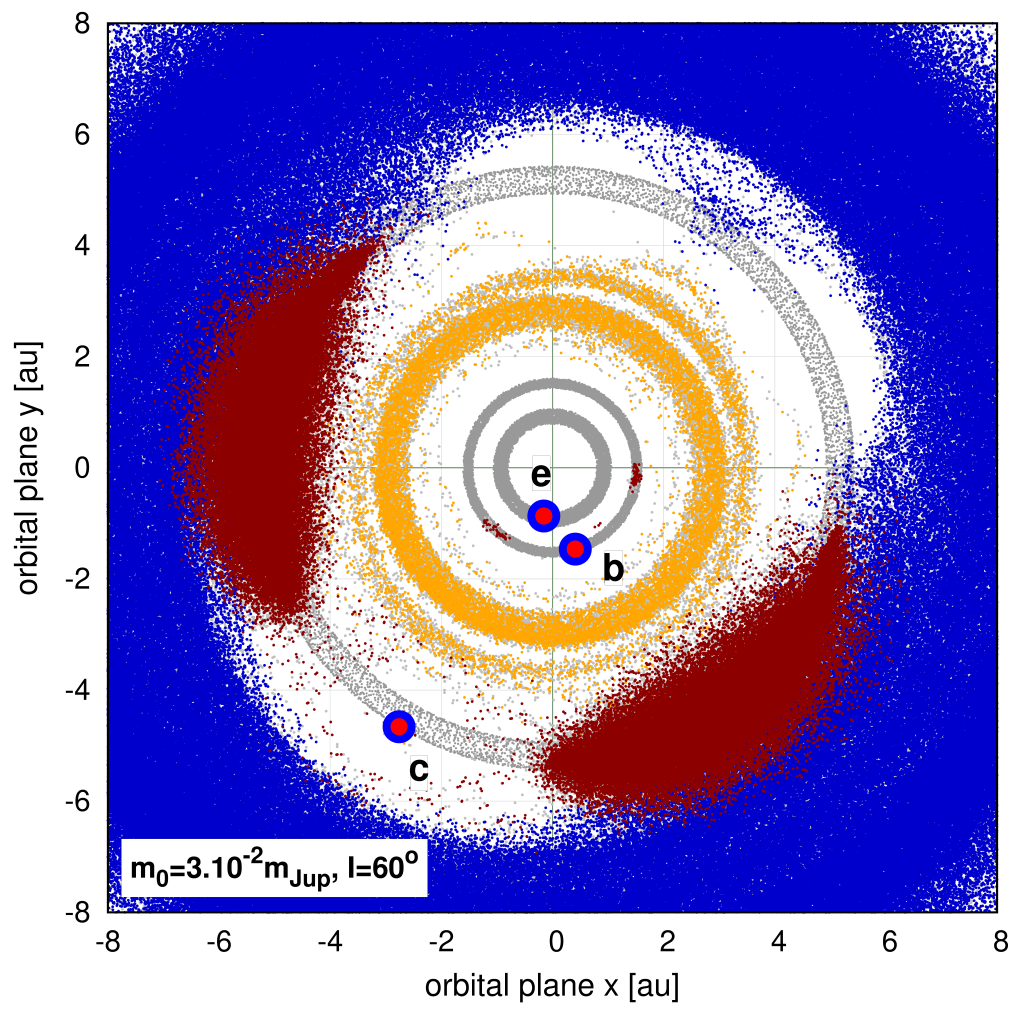}}%{../Maps/YEarth55.png}
}
\hbox{
\hbox{\includegraphics[width=0.49\textwidth]{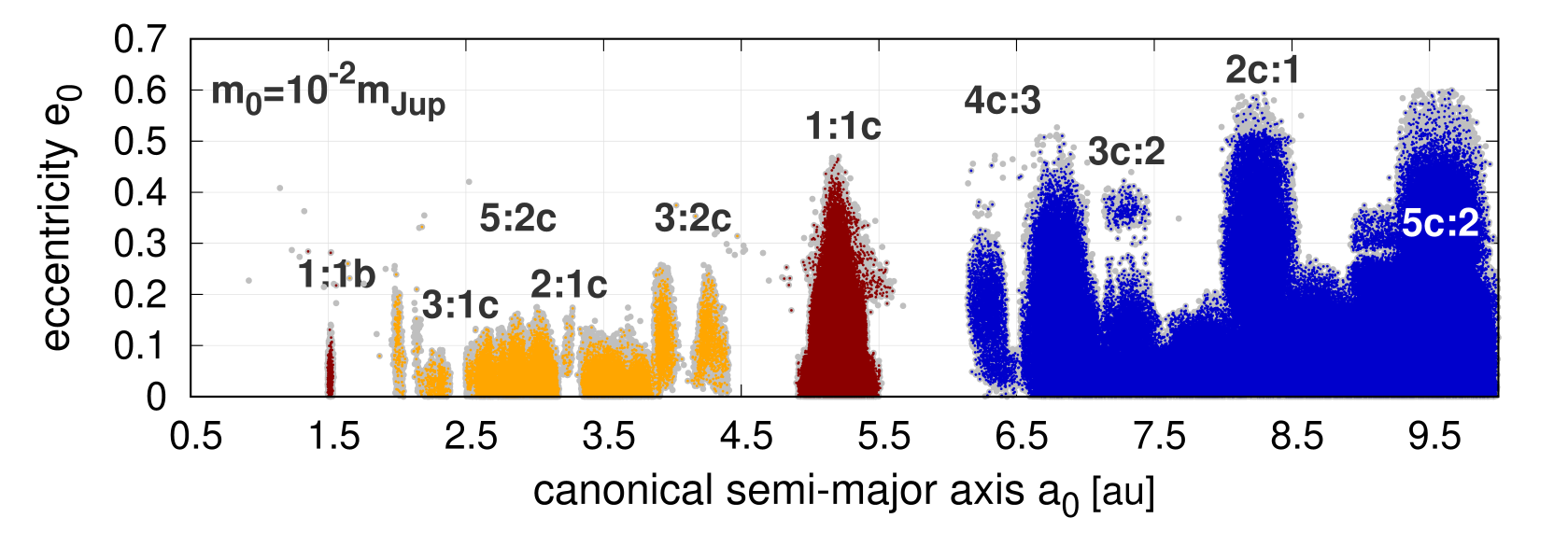}}%{../disks/DiskXY.png}
\hbox{\includegraphics[width=0.49\textwidth]{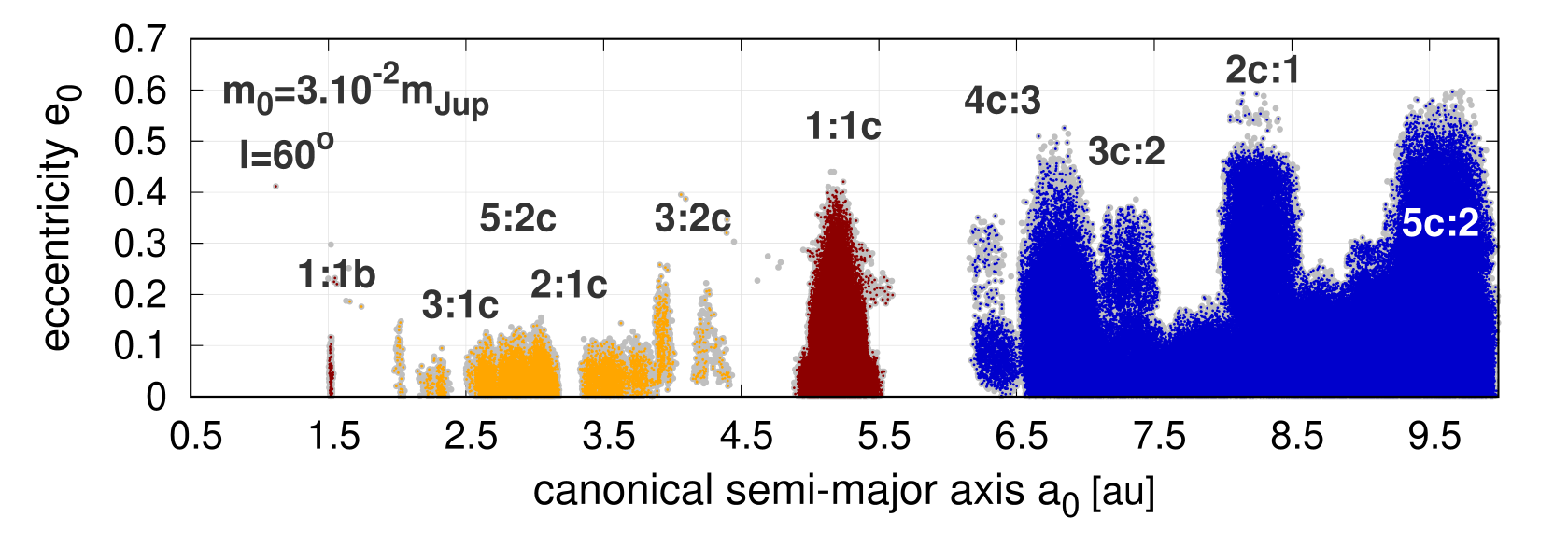}}%{../disks/DiskAE.png}
}
}
}
\caption{ 
Stable orbits in the \muarae{} system revealed by $\simeq 10^6$ solutions with $|\Ym-2|<0.007$ gathered in the $\Ym$-disk simulation for a test planet masses $m_0$ injected into the system of three outer planets with random elements, $a_0 \in [0.9,10]$~au and $e_0 \in [0,0.6])$. They are illustrated as a snapshot of astrocentric coordinates $(x, y)$ at the initial epoch $t_0$. Colors of the particles correspond to their dynamical status marked also in the panel with orbital elements, below. The initial positions of the planets are marked with filled circles. Gray rings illustrate their orbits integrated in a separate run for 0.2~Myr.
{\em The left column} is for $m_0=0.01\mJ$ (3~Earth-masses) and {\em the right column} is for $m_0=3\times 10^{-2}\mJ$ (10~masses of Earth). The integration time with the \gbs{} integrator of each initial condition is $10^5$~yr ($\simeq 10^4 \times \Pc$). 
{\em Bottom panels} are for the orbital structure of the stable orbits in terms of canonical Jacobi elements in the $(a_0,e_0)$-plane. Some two-body, lowest order MMRs with the planets are labeled, and stable orbits in their regions are marked with different colors, consistent with a snapshot of these stable solutions in the top panels.
}
\label{fig:fig14}
\end{figure*}

%
%---------------------------------------------------------------------------------------------
\subsection{Hypothetical asteroidal belts}
\label{sec:belts}
% ---------------------------------------------------------------------------------------------
%
The results for small-mass asteroids are illustrated in Fig.~\ref{fig:fig13}. Cartesian coordinates in the orbital plane of the system shown in the top-right panel are accompanied by plots for canonical elements of the test particles. {We gathered $\simeq 10^6$ stable solutions with $|\Ym-2|<0.007$ for this case}. The probe particles are marked with different colors, depending on their dynamical status: brown dots are for objects involved in 1:1c~MMR with the outermost planet \muarae{}c; orange dots are for stable orbits between \muarae{}b and \muarae{}c, and blue dots are for the Kuiper belt--like zone beyond the outermost planet. 

The edges of the debris disk formed in these regions are highly asymmetric. Also, their  non-random distribution in the plane of the osculating elements $(a_0,e_0)$ is shown in the bottom-right panel in Fig.~\ref{fig:fig13}. It was constructed based on the canonical elements determined in the Jacobi reference frame. The use of canonical elements is necessary to avoid the ``blurring'' of the distribution that would otherwise occur with astrocentric elements. In this diagram, we marked the asteroids with the same colors as in the snapshot in the orbital plane, and some of the their lowest-order MMRs with planet \muarae{}c are labeled. 

The results for the $\Ym$-model may be confronted with dynamical maps computed in the $(a_0,e_0)$-plane for fixed orbital phases of the Vesta-like particles, shown in the left column in Fig.~\ref{fig:fig13}. The maps show the phase structure in three distance regions: between the innermost pair of planet~\muarae{}d and \muarae{}e (the top panel), in the Main Belt zone (middle panel) and in the outer, Kuiper belt beyond the outermost planet \muarae{}c (bottom panel). The resonant structure of the debris disks is also clear, especially in the bottom-left map. However, as expected, the Main Belt disk structures in the two-dimensional dynamical maps are much more narrow than their representation in the $\Ym$-model, due to fixed orbital phase of the test particles.

\subsection{Earth-like planets and the habitable zone}

Although we considered low-mass asteroids in this test, stable regions can potentially host larger planets as well, in the Earth mass range. As the mass of the probing objects increases, the regions may decrease in size, both in the coordinate-- and orbital element-- planes. {This is illustrated in Fig.~\ref{fig:fig14} for Earth-mass objects (the left column) and super-Earths (the right column), respectively. That case we should interpret in terms of a potential location of the small planets rather than a representation of a physical debris disk. }

The distribution of Earth objects is very similar to the experiment for Vesta-type asteroids, as could be predicted from the similarity of this system to the restricted problem (with zero-mass asteroids). For more massive super-Earth ``asteroids'' and the inclinations of the system $I=60^{\circ}$ the stable zones shrink considerably, but the overall disks structure is still preserved. We can conclude that the $\Ym$-model scales for several orders of magnitude of the probe masses.

The results are therefore universal in the sense that we can predict the locations of e.g., Earth-like planets that are below the current detection limits. {It turns out that such small planets could be found in the habitable zone, despite \araeE{} and \araeB{} prevent stable orbits of terrestrial planets unless they are placed beyond roughly 2~au (see Fig.~\ref{fig:fig14}), or interior to $0.3$--$0.4$~au. }

Given the luminosity of \muArae{}  $L=1.9\,L_{\odot}$ and the spectral temperature $T=5820$~K \citep{Soriano2010}, the outer limiting distance roughly correspond to the orbit of Mars in the Solar system. Indeed, for an Earth-like planet, the inner radius of the runaway greenhouse effect is $r_{\rm min}=1.31$~au, the radius of maximum greenhouse effect $r_{\rm max}=2.30$~au, and the radius for early Mars zone $r_{\rm EM}=2.42$~au \citep[][their habitable zone calculator]{Kopparapu2014}. Therefore, habitable Earth-like planets could be found in a small region of Lagrangian (Trojan) 1:1b orbits around \muarae{}b as well on the inner edge of the Main Belt, up to the 3:1c MMR gap (see the elements distribution in Fig.~\ref{fig:fig14}).

%_______________________________________________________________________________
%
\section{Conclusions}
\label{sec:conclusions}
%_______________________________________________________________________________
%

The \muarae{} planetary system is one of the first detected multi-planet configurations with a mass-diverse planets, and it comprises of a warm Neptune, a Saturn-mass planet, and two massive Jupiter-mass objects. The precision RV data available in public archives, spanning at least 1.5~outermost periods, makes it already possible to tightly constrain the orbits and minimal masses of the planetary companions to $1\sigma \simeq 0.02\,\mJ$.  Unfortunately, given a low accuracy of the HST astrometry reported in \citep{Benedict2022}, and insufficient detection limits (estimated here independently), we restricted the analysis to the RV data only.

We improved kinematic (Keplerian) models reported more than 15~years ago \citep{Gozdziewski2007a,Pepe2007}, as well as in the very recent paper by \cite{Benedict2022}. Our Newtonian RV models of the \muarae{} system imply its long-term stable, Solar system-like orbital architecture. The planets revolve in low-eccentricity orbits determined with significantly reduced uncertainties $\simeq 0.01$ w.r.t. the prior literature, closely resembling the Earth--Mars--Jupiter sequence. Other orbital elements, and particularly the semi-major axes are bounded $0.02$~au for the outermost planet, and to just $0.001$--$0.002$~au for remaining inner massive companions. Limiting uncertainty of the outermost semi-major axis to $\simeq 27$~days means a qualitative improvement, compared to uncertainties of $700$--$1300$~days reported in \citep{Gozdziewski2007a} and \citep{Pepe2007}.  

Using the dynamical maps technique, we found that the nominal \ics{} cover regions in the phase space within several $\sigma$ error bars that correspond to long-term stable evolution. The direct numerical integrations indicate stable orbital evolution of the best-fitting models for at least 6.7~Gyr (i.e., the lifetime of the star).

The present RV data do not make it possible to fully constrain the system inclination. However, it does not influence the stability in a wide range between $90^{\circ}$ and $\simeq 20^{\circ}$. In this range, coplanar systems remain in similarly wide and safe zones of stable motions, despite of planet masses enlarged a few times, in accord with the $m \sin I$ relation.   Moreover, we found a close overlap of the dynamical stability with the best-fitting models in the sense that there is clear maximum of the posterior distribution for $\lnL$ and a steep increase of the RMS at $I \simeq 20^{\circ}$--$30^{\circ}$. This means that all the masses would remain certainly below the brown dwarf mass range. It also proves that the analysed RV data bring information on the mutual interactions between the system components.
 
The meaningfully constrained orbits make it possible to globally investigate the global dynamical structure of the system.  The inner pair of Saturn-Jupiter--mass planets is close to the 2e:1b~MMR, but is significantly and {systematically} separated from this resonance. Similarly, the outer pair is close to the 6b:1c~MMR but also is meaningfully far from it. This result may be important since it adds a new observational evidence on a near-resonant, well characterised multiple system with Jovian-mass planets. 

Multiple planetary systems, especially in the lower mass range detected by the Kepler mission, exhibit excess of planets close to first-order MMR (2:1 and 3:2), with the period ratio slightly higher than the resonant value \citep[e.g.][and references therein]{Petrovich2013,Ramos2017,Delisle2014,Marzari2018}. There is a debate in the literature about the origin of this effect.  It has recently been shown \citep{Marzari2018} that the presence of a massive circumbinary disk can significantly affect the resonant behavior of a pair of planets, shifting the resonant position and reducing the size of the stability region. Dissipation of the disk may explain some exosystems that are close to the MMR but not trapped within it. If such mechanism was active in the \muArae{} system, the current, near 2e:1b MMR for the inner pair could be a signature of a massive circumstellar disk in the past and its remnants in the form of asteroid belts at present. In this context, the evolution of the \muArae{} system serves a particularly interesting scenario. The near-resonant pair is accompanied by a more distant high-mass companion, also near higher-order 6c:1b~MMR of the outer pair, which certainly enriches the dynamical setup.

The orbital architecture permits for the presence of massive debris disks, indeed, as they might survive between the planets. There is especially wide region between the outer pair, spanning the semi-major axes range of $(1.5,5.2)$~au; also there is such a vastly wide stable region beyond the outermost planet, starting at $\simeq 6.5$\,\au{} and huge Trojan islands coorbital with the outermost planet. Simulations of these debris disks reveal their strongly resonant structure {that is preserved in a wide range of probe masses, between Vesta-like asteroids and super-Earths with $\simeq 10$~Earth masses.} Te debris disks would be (obviously) strongly influenced by the MMRs with the Jovian planets. Their short-term MMR structure closely resembles the Main Belt and the Kuiper Belt in the Solar system. 

Prospects to detect relatively massive, super-Earth--mass objects in the zone around $0.3$~au--$0.5$~au or in other parts of the system, where stable orbits of are possible, are uncertain but unlikely. The semi-amplitude of their RV signals would be comparable with the intrinsic stellar jitter variability. We did not detect significant periods in the residuals of the RV models other than those identified with the known planets.

Because \muArae{} has a fairly large parallax ($\simeq 65$~mas), it may be an interesting and promising target for \alma{} and other instruments to detect dust emissions, and set additional limits on the presence of small planets in outer parts of the system. In addition, the detection of debris disks, especially the outer one, can help better constrain the inclination of the system. 

Monitoring the RV variability of the star still seems plausible, as it may permit to characterise the system even better, once the \gaia{} DR4 catalogue is released. Our simulation of the \iad{} measurements with the help of \htof{} package \citep{Brandt2021} reveal that the two outer planets will be astrometrically detectable with very high S$/$N, provided the uncertainty of the \iad{} time series on the level of $0.1$~mas. Moreover, we have shown that the mutual gravitational interactions can be detected in the RV data up to the middle of 2015. Additional precision RV observations might greatly help to break or reduce the $m\sin I$ degeneracy, and confirm or rule out the inclination of the system $I \simeq 20^{\circ}$--$30^{\circ}$ indicated by our Bayesian MCMC sampling experiments.

Finally, the highly hierarchical configuration of \muArae{} is a new test-bed for our new fast indicator \rem{} \citep{Panichi2017} that helps to analyse the structure of the phase space in terms of the most accurate, Newtonian representation of the data. Despite analytical approximations for the motion of the innermost planet may be constructed \citep{Farago2009}, the simple \rem{} algorithm based on the canonical leap-frog scheme offers a sufficient numerical efficiency to derive the dynamical maps through integrating the exact equations of motion of the whole system. It is also fully compatible with more CPU demanding \megno{} technique, especially for systems in regions of the the phase space which are filled with mostly stable solutions.
%
%_____________________________________________________________________________
%
\section{Acknowledgements}
%_______________________________________________________________________________
%
We thank the anonymous reviewer for critical, constructive and very helpful comments that greatly improved this work. We thank Dr Franz Benedict for providing the RV data set for \muarae{} prior to publication. We are very grateful to Karolina B\c{a}kowska,  Agnieszka S\l{}owikowska and Pawe\l{} Zieli\'nski for help and a discussion regarding photometry and RVs of \muArae{}. We thank the Pozna\'n Supercomputer and Network Centre (PCSS, Poland) for computing resources (grant No. 529) and the long-term, generous support.
%_______________________________________________________________________________
%

\section{Data availability}
The Radial Velocity time series referenced in this paper as data set \Done{}
are available in their source form, as published by \cite{Benedict2022} and as
data sets \Dtwo{} and \Dthree{} from \cite[][\url{
https://doi.org/10.1051/0004-6361/201936686}]{Trifonov2020}, also
\url{https://github.com/3fon3fonov/HARPS_RVBank}. All other data presented
in Tables~\ref{tab:tab1}, \ref{tab:tab2} and Figures,
underlying this article will be shared on reasonable request to the corresponding author.

\bibliographystyle{mn2e}
\bibliography{ms}
\label{lastpage}
%\end{document}

\newpage
\setcounter{figure}{0}
\setcounter{table}{0}
\renewcommand{\thefigure}{A\arabic{figure}}
\renewcommand{\thetable}{A\arabic{table}}

%\smallskip
\section*{On-line Supplemntary Material}
The following Section contains supplementary MCMC corner plots illustrating posterior probability distribution for three
RVs models investigated in the paper and the numerical initial conditions to
reproduce some figures in the paper
%
%
% \begin{figure*}
% \centerline{
% \vbox{
% \hbox{
% \includegraphics[width=0.5\textwidth]{figA1a}%{../Doppler7.Mcmc/muAraP1.pdf}
% \includegraphics[width=0.5\textwidth]{figA1b}%{../Doppler7.Mcmc/muAraP4.pdf}
% }
% \hbox{
% \includegraphics[width=0.5\textwidth]{figA1c}%{../run777777/muAraP1.pdf}
% \includegraphics[width=0.5\textwidth]{figA1d}%{../run777777/muAraP4.pdf}
% }
% }
% }
% \caption{
% One-- and two--dimensional projections of the posterior probability distribution for orbital parameters of the innermost (left column) and the outermost (the right column) planet, respectively. The top row is for the Keplerian model, and the bottom row is for the Newtonian model. The parameters are expressed in units consistent with Table~\ref{tab:tab1}. The semi-amplitude $K_i$ is equivalent to the mass $m_i$, and the orbital period $P_i$ is equivalent to the semi-major-axis $a_i$. The MCMC chain length {is 128,000 iterations in each of 384 different instances selected in a small ball around a best-fitting solution found with the evolutionary algorithms}. Parameter uncertainties are estimated as 16th and 84th percentile samples around the median values at 50th percentile. 
% }
% \label{fig:figA1}
% \end{figure*}

\bigskip

\begin{figure*}
\centerline{
\vbox{
\hbox{
\includegraphics[width=0.5\textwidth]{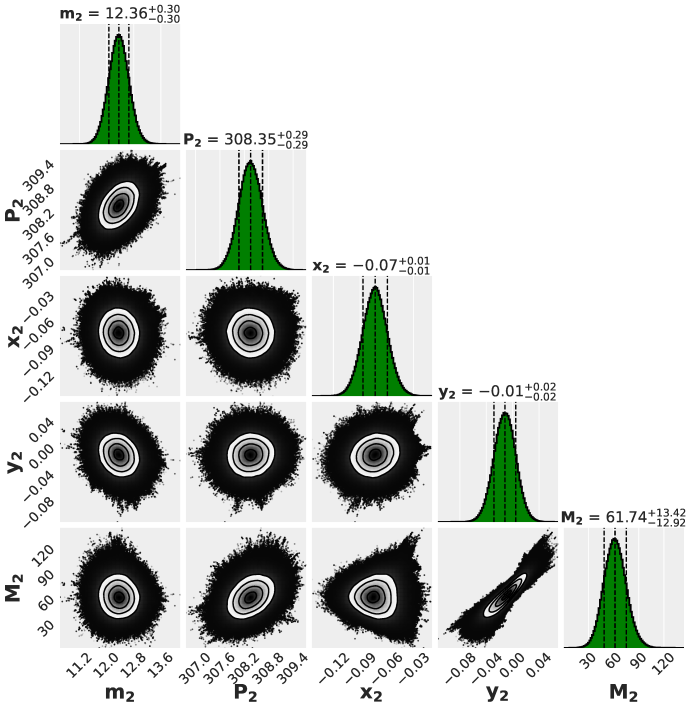}%{../Doppler7.Mcmc/muAraP2.pdf}
\includegraphics[width=0.5\textwidth]{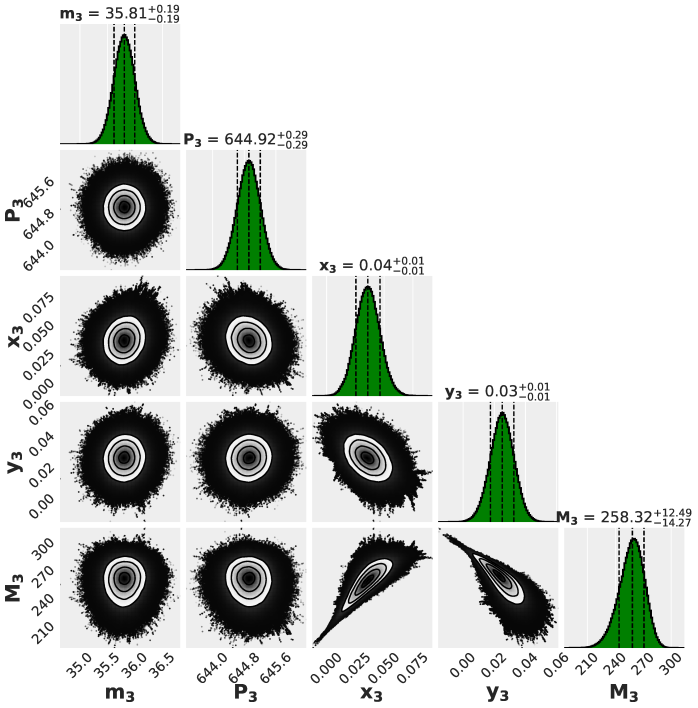}%{../Doppler7.Mcmc/muAraP3.pdf}
}
\hbox{
\includegraphics[width=0.5\textwidth]{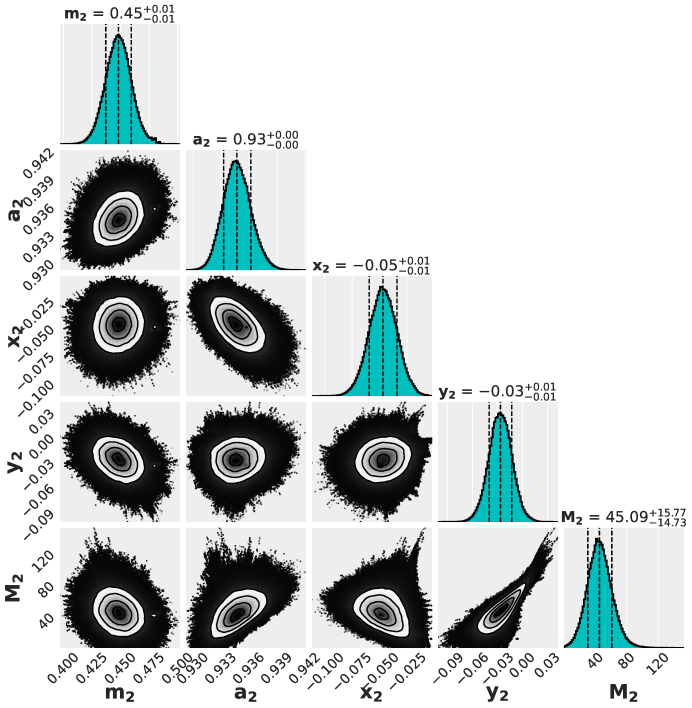}%{../run777777/muAraP2.pdf}
\includegraphics[width=0.5\textwidth]{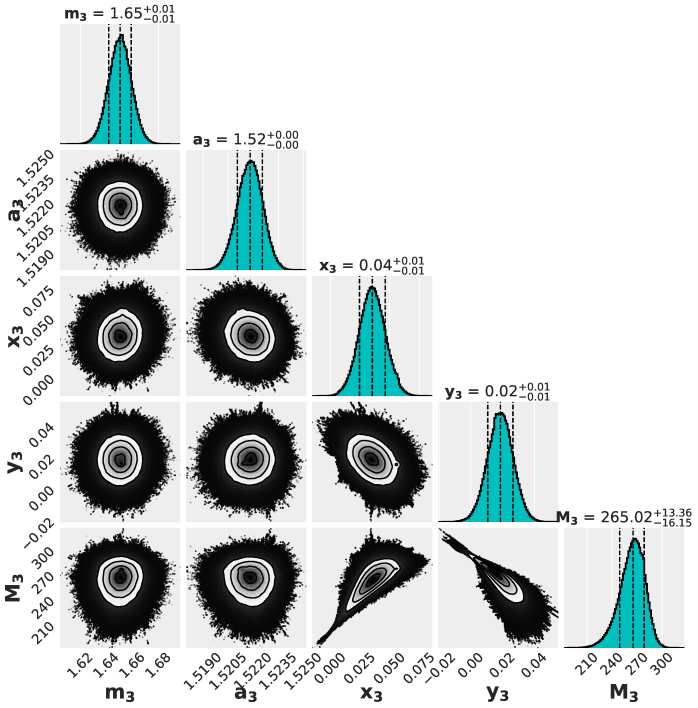}%{../run777777/muAraP3.pdf}
}
}
}
\caption{
One-- and two--dimensional projections of the posterior probability distribution for orbital parameters of the innermost (left column) and the outermost (the right column) planet, respectively. The top row is for the Keplerian model, and the bottom row is for the Newtonian model. The parameters are expressed in units consistent with Table~2.
The semi-amplitude $K_i$ is equivalent to the mass $m_i$, and the orbital period $P_i$ is equivalent to the semi-major-axis $a_i$. The MCMC chain length is 128,000 iterations in each of 384 different instances selected in a small ball around a best-fitting solution found with the evolutionary algorithms. Parameter uncertainties are estimated as 16th, and 84th percentile samples around the median values (50th percentile).
}
\label{fig:figA1}
\end{figure*}

\begin{figure*}
\centerline{
\hbox{
\includegraphics[width=0.9\textwidth]{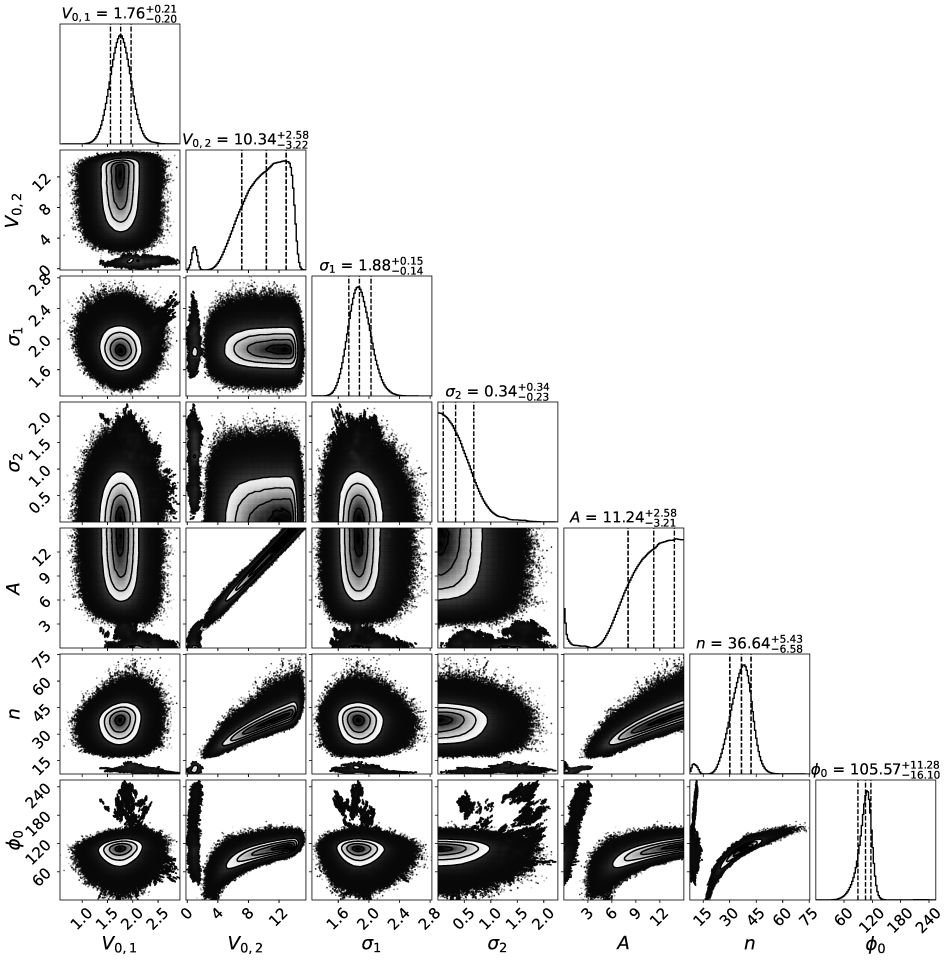}%{../Doppler5.Mcmc/muAraeEjit.pdf}
}
}
\caption{
A fragment of the corner plot for posterior samples  for the Keplerian model with an instrumental drift attributed to \ucles{} measurements. Offsets $V_{0,1}$, $V_{0,2}$ and jitters $\sigma_1$ and $\sigma_2$ for \harps{} and \ucles{}, respectively, as well as the semi-amplitude $A$ of the drift signal are expressed in \ms{}. The $n$ parameter is the drift frequency converted to and expressed in years, and the phase shift of the drift $\phi_0$ is expressed in degrees.
}
\label{fig:figA2}
\end{figure*}

\begin{verbatim}
Astrocentric Keplerian Elements to reproduce Fig. 11 (bottom panel) and Fig. 12

Stellar mass      1.13002842 Solar masses     

# m[mJup]     a[au]        e     Inc[deg]  Om[deg]  om[deg]       M[deg]
#
# Planet d
0.0922327   0.0923196  0.1369002   19.0    0.0     -166.3286582  144.888 7559
# Plabet e 
1.3473597   0.9290769  0.0816879   19.0    0.0     -173.2692105  153.9525959
# Planet b
4.9087852   1.5262249  0.0472992   19.0    0.0       22.3406827  234.2662176
# Planet c
5.8499813   5.1882224  0.0363553   19.0    0.0       73.8847971   86.8947565
\end{verbatim}

\end{document}